\documentclass{article}
\usepackage{theorem,amsmath,amssymb,graphicx}
\newtheorem{Th}{Theorem}[section]
\newtheorem{Le}{Lemma}[section]
\newtheorem{Co}{Corollary}[section]

\newtheorem{Rem}{Remark}[section]
\newcommand{\ol}{\overline}

\begin{document}
\renewcommand{\theequation}{\arabic{section}.\arabic{equation}}
\title{Mixed boundary value problems for the Navier-Stokes system in polyhedral domains}
\date{}
\author{by V.~Mazya and J.~Rossmann}
\maketitle
\begin{abstract}
Mixed boundary value problems for the Navier-Stokes system in a polyhedral domain are considered.
Different boundary conditions (in particular, Dirichlet, Neumann, slip conditions)
are prescribed on the faces of the polyhedron. The authors obtain regularity results for
weak solutions in weighted (and non-weighted) $L_p$ Sobolev and H\"older spaces. \\


\end{abstract}

\maketitle

\setcounter{section}{-1}
\section{Introduction}
Steady-state flows of incompressible viscous Newtonian fluids are
modeled  by the Navier-Stokes equations
\begin{equation} \label{Navier-S}
-\nu \, \Delta u + (u\cdot \nabla)\, u + \nabla p = f, \qquad
\nabla\cdot u = g
\end{equation}
for the velocity $u$ and the pressure $p$. For this system, one
can consider different boundary conditions. For example on solid
walls, we have the Dirichlet condition $u=0$. On other parts of
the boundary (an artificial boundary such as the exit of a
channel, or a free surface) a no-friction condition (Neumann
condition) $2\nu \varepsilon(u)\, n -pn =0$ may be useful. Here
$\varepsilon(u)$ denotes the matrix with the components $\frac 12
(\partial_{x_i}u_j+\partial_{x_j}u_i)$, and $n$ is the outward
normal. Note that the Neumann problem naturally appears in the
theory of hydrodynamic potentials (see \cite{Lad}). It is also of
interest to consider boundary conditions containing components of
the velocity and of the friction. Frequently used combinations are
the normal component of the velocity and the tangential component
of the friction (slip condition for uncovered fluid surfaces) or
the tangential component of the velocity and the normal component
of the friction (condition for in/out-stream surfaces).

In the present paper, we consider  mixed boundary value problems
for the system (\ref{Navier-S}) in a three-dimensional domain
${\cal G}$ of polyhedral type, where components of the velocity
and/or the friction are given on the boundary. To be more precise,
we have one of the following boundary conditions on each face
$\Gamma_j$:
\begin{itemize}
\item[(i)] $u=h$,
\item[(ii)] $u_\tau=h,\quad -p+2\varepsilon_{n,n}(u) = \phi$,
\item[(iii)] $u_n  = h, \quad \varepsilon_{n,\tau}(u)=\phi$,
\item[(iv)] $-p n + 2\varepsilon_n (u) = \phi$,
\end{itemize}
where $u_n=u\cdot n$ denotes the normal and $u_\tau=u-u_n n$ the
tangential component of $u$, $\varepsilon_n(u)$ is the vector
$\varepsilon(u)\, n$, $\varepsilon_{n,n}(u)$ is the normal
component and $\varepsilon_{n,\tau}(u)$ the tangential component
of $\varepsilon_n(u)$.

Weak solutions, i.e. variational solutions $(u,p) \in
W^{1,2}({\cal G})^3 \times L_2({\cal G})$, always exist if the
data are sufficiently small. In the case when the boundary
conditions (ii) and (iv) disappear, such solutions exist for
arbitrary $f$ (see the books by Ladyzhenskaya \cite{Lad}, Temam
\cite{Temam}, Girault and Raviart \cite{Girault}). Our goal is to
prove regularity assertions for weak solutions. As is well-known,
the local regularity result
\[ (u,p) \in W^{l,s} \times W^{l-1,s} \]
is valid outside an arbitrarily small neighborhood of the edges
and vertices if the data are sufficiently smooth. Here $W^{l,s}$
denotes the Sobolev space of functions which belong to $L_s$
together with all derivatives up to order $l$. The same result
holds for the H\"older space $C^{l,\sigma}$. Since solutions of
elliptic boundary value problems in general have singularities
near singular boundary points, the result cannot be globally true
in ${\cal G}$ without any restrictions on $l$ and $s$. Here we
give a few particular regularity results which are consequences of
more general theorems proved in the present paper. Suppose that
the data belong to corresponding Sobolev or H\"older spaces and
satisfy certain compatibility conditions on the edges. Then the
following smoothness of the weak solution is guaranteed and is the
best possible.
\begin{itemize}
\item If $(u,p)$ is a solution of the Dirichlet problem in an arbitrary polyhedron or a solution of
the Neumann problem in an arbitrary Lipschitz graph polyhedron,
then
\begin{eqnarray*}
&& (u,p) \in W^{1,3+\varepsilon}({\cal G})^3 \times L_{3+\varepsilon}({\cal G}), \\
&& (u,p) \in W^{2,4/3+\varepsilon}({\cal G})^3 \times W^{1,4/3+\varepsilon}({\cal G}), \\
&& u \in C^{0,\varepsilon}({\cal G})^3.
\end{eqnarray*}
Here $\varepsilon$ is a positive number depending on the domain
${\cal G}$.
\item If $(u,p)$ is a solution of the Dirichlet problem in a convex polyhedron, then
\begin{eqnarray*}
&& (u,p) \in W^{1,s}({\cal G})^3 \times L_s({\cal G}) \quad \mbox{ for all } s, \ 1< s < \infty, \\
&& (u,p) \in W^{2,2+\varepsilon}({\cal G})^3 \times W^{1,2+\varepsilon}({\cal G}), \\
&& (u,p) \in C^{1,\varepsilon}({\cal G})^3 \times
C^{0,\varepsilon}({\cal G}).
\end{eqnarray*}
\item If $(u,p)$ is a solution of the mixed problem in an arbitrary polyhedron with the
Dirichlet and Neumann boundary condition prescribed arbitrarily on
different faces, then
\[
(u,p) \in W^{2,8/7+\varepsilon}({\cal G})^3 \times
W^{1,8/7+\varepsilon}({\cal G}).
\]
\item Let $(u,p)$ be a solution of the mixed boundary value problem with slip condition (iii)
on one face $\Gamma_1$ and Dirichlet condition on the other faces.
Then
\begin{eqnarray*}
&& (u,p) \in W^{1,3+\varepsilon}({\cal G})^3 \times L_{3+\varepsilon}({\cal G}) \quad\mbox{if $\theta< \frac 32 \pi$},\\
&& (u,p) \in W^{2,2+\varepsilon}({\cal G})^3 \times
W^{2,2+\varepsilon}({\cal G})  \quad
  \mbox{if ${\cal G}$ is convex and $\theta< \pi/2$}, \\
&& (u,p) \in C^{1,\varepsilon}({\cal G})^3 \times
C^{0,\varepsilon}({\cal G})  \quad
  \mbox{if ${\cal G}$ is convex and   $\theta< \pi/2$},
\end{eqnarray*}
where $\theta$ is the maximal angle between $\Gamma_1$ and the
adjoining faces.
\end{itemize}
General facts of such a kind imply various precise regularity
statements for special domains. Let us consider for example the
flow outside a regular polyhedron $G$. On the boundary of $G$, the
Dirichlet condition is prescribed. Then we obtain $(u,p) \in
W^{2,s} \times W^{1,s}$ on every bounded subdomain of the
complement of $G$, where the best possible value for $s$ is
(with all digits shown correct) \\
\begin{figure}[h]
\includegraphics{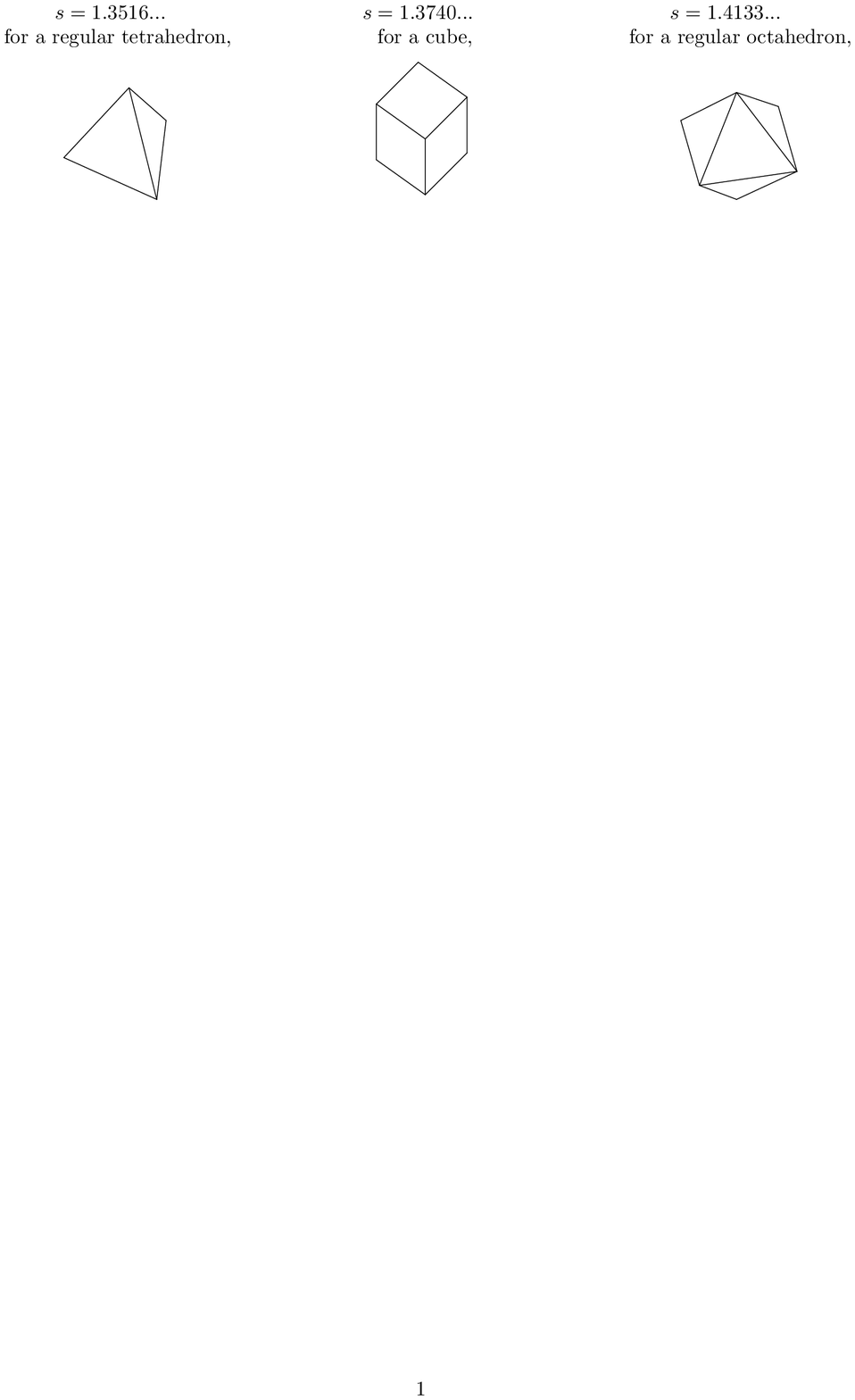}
\end{figure}

\newpage
\begin{figure}[h]
\includegraphics{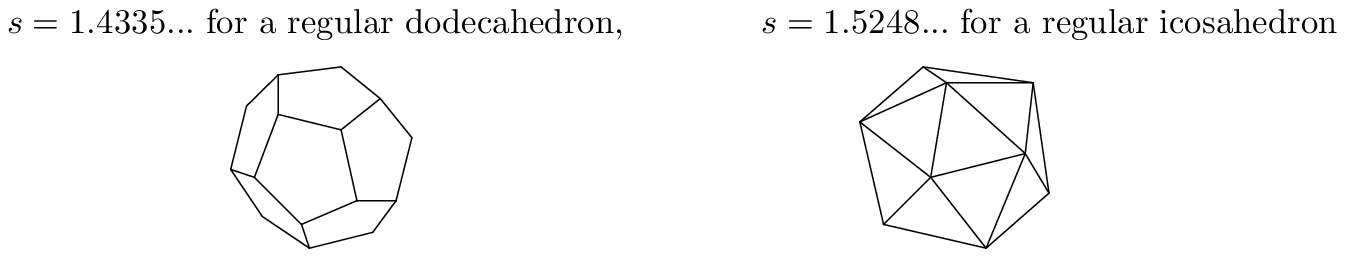}
\end{figure}

In the last decades, numerous papers appeared which treat boundary
value problems for elliptic equations and systems in piecewise
smooth domains. For the stationary linear Stokes system see e.g.
the  references in the book \cite{kmr2}. The properties of
solutions of the Dirichlet problem to the nonlinear Navier-Stokes
system in 2-dimensional polygonal domains were studied in papers
by Kondrat'ev \cite{kon-67}, Kellogg and Osborn \cite{Kellogg},
Kalex \cite{Kalex}, Orlt and S\"andig \cite{Orlt}. In particular,
Kellogg and Osborn proved that the solution of the Dirichlet
problem belongs to $W^{2,2}({\cal G})^3 \times W^{1,2}({\cal G})$
if $f \in L_2({\cal G})$ and the polygon ${\cal G}$ is convex.
Kalex, Orlt and S\"andig  considered solutions of mixed boundary
value problems in polygonal domains. Mixed boundary value problems
with boundary conditions (i) and (iii) in 3-dimensional domains
with smooth non-intersecting edges were handled by Solonnikov
\cite{Solonn-83}, Maz'ya, Plamenevski\u{\i} and Stupyalis
\cite{mps-84}. They proved in particular the solvability in
weighted Sobolev and H\"older spaces. For the case of the
Dirichlet problem and a polyhedral domain, solvability and
regularity results in weighted Sobolev and H\"older spaces were
proved by Maz'ya and Plamenevski\u{\i} \cite{mp-83}. Concerning
regularity results in $L_2$ Sobolev spaces, we refer also to the
papers of Nicaise \cite{Nicaise-97} (Dirichlet problem), Ebmeyer
and  Frehse \cite{Ebmeyer-01} (mixed problem  with boundary
conditions (i) and (iii)). Ebmeyer and Frehse proved that $(u,p)
\in W^{s,2}({\cal G})^3 \times W^{s-1,2}({\cal G})$ with arbitrary
real $s<3/2$ if the angle at the edge, where the boundary
conditions change, is less than $\pi$. Finally, we mention the
papers by Deuring and von Wahl \cite{Deuring}, Dindos and Mitrea
\cite{Dindos} dealing with the Navier-Stokes system in Lipschitz
domains.

The present paper consists of four sections. Section 1 concerns
the existence of weak solutions in $W^{1,2}({\cal G})^3\times
L_2({\cal G})$. In Section 2, we introduce and study weighted
Sobolev and H\"older space. Here the weights are powers of the
distances $\rho_j$ and $r_k$ to the vertices and edges of the
domain ${\cal G}$, respectively. In particular, we establish
imbedding theorems for these spaces. In contrast to the papers
\cite{mp-83,mps-84,Solonn-83}, we use weighted spaces with
``nonhomogeneous'' norms. The weighted Sobolev space
$W_{\beta,\delta}^{l,s}({\cal G})$ in our paper is defined as the
set of all functions $u$ in ${\cal G}$ such that
\[
\prod_j \rho_j^{\beta_j-l+|\alpha|} \ \prod_k \Big(
\frac{r_k}{\rho} \Big)^{\delta_k} \  \partial_x^\alpha u\in
L_s({\cal G})
\]
for all $|\alpha|\le l$, where $\rho=\min_j \rho_j$. The norm in
the weighted H\"older space $C_{\beta,\delta}^{l,\sigma} ({\cal
G})$ has a similar structure. The use of weighted spaces with
nonhomogeneous norms has several advantages. First, these spaces
are applicable to a wider class of boundary value problems. For
$\beta=0$ and $\delta=0$ they are closely related to the
nonweighted spaces. Furthermore in some cases (e.g. the Dirichlet
problem when the edge angles are less than $\pi$), it is possible
to obtain higher regularity results when considering solutions in
weighted spaces with nonhomogeneous norms. So we can partially
improve the results in \cite{mp-83,mps-84,Solonn-83}. The main
results of the paper are contained in Section 3. For the proofs,
we use results of our previous papers \cite{mr-04b,mr-05}, where
we studied mixed boundary value problems for the linear Stokes
system. We show in particular that the weak solution $(u,p)$
belongs to the weighted space $W_{\beta,\delta}^{2,s}({\cal
G})^3\times W_{\beta,\delta}^{1,s}({\cal G})$ if the data are from
corresponding spaces, satisfy certain compatibility conditions,
and the numbers $\frac 52 -\beta_j-\frac 3s$ and $2-\delta_k-\frac
2s$ are positive and sufficiently small. The precise conditions on
$\beta$ and $\delta$ are given in terms of eigenvalues of certain
operator pencils. The general results in Section 3 together with
estimates for the eigenvalue of these pencils (see \cite{kmr2})
allow us in particular to deduce regularity assertions in
nonweighted Sobolev and H\"older spaces. A number of examples is
given at the end of Section 3.

The last section concerns the solvability in the space
$W_{\beta,\delta}^{1,s}({\cal G})^3\times
W_{\beta,\delta}^{0,s}({\cal G})$, where $s$ may be less than 2.
Here, we assume that the Dirichlet condition is given on at least
one of the adjoining faces of every edge $M_k$. One of our results
is the following. Let ${\cal G}$ be an arbitrary polyhedron. Then
the problem (\ref{Navier-S}) with $g=0$ and Dirichlet condition
$u=0$ on the boundary has a weak solution $(u,p)\in W^{1,s}({\cal
G})^3 \times L_s({\cal G})$ for arbitrary $f\in W^{-1,s}({\cal
G})^3$, $3/2<s<3$, provided the norm of $f$ is sufficiently small.
The same result holds for the mixed problem with boundary
condition (i)--(iii) if we suppose that the angles at the edges
where the boundary conditions change are less or equal to
$3\pi/2$.

\setcounter{section}{0} \setcounter{equation}{0}
\section{Weak solutions of the boundary value problem}

\subsection{The domain}

In the following, let ${\cal D}$ be the dihedron
\begin{equation} \label{dih}
\{ x=(x_1,x_2,x_3)\in {\Bbb R}^3:\ 0< r < \infty,\ -\theta/2<
\varphi <\theta/2,\ x_3\in {\Bbb R}\},
\end{equation}
where $r,\varphi$ are the polar coordinates in the
$(x_1,x_2)$-plane, $r=(x_1^2+x_2^2)^{1/2}$, $\tan
\varphi=x_2/x_1$. Furthermore, let ${\cal K} = \{ x\in {\Bbb
R}^3:\ x/|x| \in \Omega\}$ be a polyhedral cone with plane faces
$\Gamma_1,\ldots,\Gamma_N$ and edges $M_1,\ldots,M_N$.

The bounded domain ${\cal G}\subset {\Bbb R}^3$ is said to be a
{\em domain of polyhedral type} if
\begin{itemize}
\item[(i)] the boundary $\partial{\cal G}$ consists of smooth (of class $C^\infty$)
  open two-dimensional manifolds $\Gamma_j$ (the faces of ${\cal G}$), $j=1,\ldots,N$,
  smooth curves $M_k$ (the edges), $k=1,\ldots,m$, and vertices $x^{(1)},\ldots,x^{(d)}$,
\item[(ii)] for every $\xi\in M_k$ there exist a neighborhood ${\cal U}_\xi$ and
  a diffeomorphism (a $C^\infty$ mapping) $\kappa_\xi$ which maps
  ${\cal G}\cap {\cal U}_\xi$ onto ${\cal D}_\xi \cap B_1$, where ${\cal D}_\xi$ is a
  dihedron of the form (\ref{dih}) and $B_1$ is the unit ball,
\item[(iii)] for every vertex $x^{(j)}$ there exist a neighborhood ${\cal U}_j$ and
  a diffeomorphism $\kappa_j$ mapping ${\cal G}\cap {\cal U}_j$ onto ${\cal K}_j \cap B_1$,
  where ${\cal K}_j$ is a polyhedral cone with vertex at the origin.
\end{itemize}
The set $M_1\cup\cdots\cup M_m\cup\{ x^{(1)},\ldots,x^{(d)}\}$ of
the singular boundary points is denoted by ${\cal S}$.

\subsection{Formulation of the problem}

For every face $\Gamma_j$, $j=1,\ldots,N$, let a number $d_j\in
\{0,1,2,3\}$ be given. We consider the boundary value problem
\begin{eqnarray} \label{NS1}
&& -\nu \Delta u + \sum_{j=1}^3 u_j \, \partial_{x_j}u + \nabla p
= f,\quad -\nabla \cdot u =g\quad
  \mbox{ in } {\cal G}, \\ \label{NS2}
&& S_j u = h_j,\quad N_j(u,p)= \phi_j\quad \mbox{on }\Gamma_j,\
j=1,\ldots,N,
\end{eqnarray}
where
\[
S_j u = \left\{ \begin{array}{ll} u & \mbox{ if }d_j=0,\\ u_\tau & \mbox{ if }d_j=1, \\
  u_n & \mbox{ if }d_j=2,\end{array}\right. \qquad
N_j(u,p) = \left\{ \begin{array}{ll} -p+2\nu \varepsilon_{nn}u &
\mbox{ if }d_j=1,\\ \varepsilon_{n\tau}(u)
  & \mbox{ if }d_j=2, \\ -pn + 2\nu \varepsilon_n(u) & \mbox{ if }d_j=3.\end{array}\right.
\]
By a weak solution of the problem (\ref{NS1}), (\ref{NS2}), we
mean a vector function $(u,p)\in W^{1,2}({\cal G})^3\times
L_2({\cal G})$ satisfying
\begin{eqnarray} \label{a1}
&& b(u,v) + \int_{\cal G} \sum_{j=1}^3 u_j\, \frac{\partial
u}{\partial x_j}\cdot v\, dx
  - \int_{\cal G} p\, \nabla\cdot v\, dx = F(v)\ \mbox{ for all }v\in V, \\ \label{a2}
&& -\nabla \cdot u =g \  \mbox{ in } {\cal G},\quad S_j u = h_j\
\mbox{ on }\Gamma_j,\ j=1,\ldots,N,
\end{eqnarray}
where $V=\{ u\in W^{1,2}({\cal G})^3:\, S_ju|_{\Gamma_j}=0,\
j=1,\ldots,N\}$,
\begin{eqnarray} \label{a3}
&& b(u,v)= 2\nu \int_{\cal G} \sum_{i,j=1}^3 \varepsilon_{i,j}(u)\, \varepsilon_{i,j}(v)\, dx,\\
&& F(v) = \int_{\cal G} (f+\nabla g)\cdot v\, dx + \sum_{j=1}^n
\int_{\Gamma_j} \phi_j\cdot v\, dx. \label{a4}
\end{eqnarray}
Note that for arbitrary $u\in W^{1,2}({\cal G})^3$, the functional
$\displaystyle v \to \int_{\cal G}  u_j\, \frac{\partial
u}{\partial x_j}\cdot v\, dx$ is continuous on $W^{1,2}({\cal
G})^3$. This follows from the inequality
\[
\Big| \int_{\cal G} u_j\, \frac{\partial u}{\partial x_j}\cdot v\,
dx\Big|
  \le \| u_j\|_{L_4({\cal G})} \ \| \partial_{x_j} u\|_{L_2({\cal G})^3} \ \| v\|_{L_4({\cal G})^3}
\]
and the continuity  of the imbedding $W^{1,2}({\cal G}) \subset
L_4({\cal G})$.

\subsection{Existence of solutions of the linearized problem}

We consider the weak solution of the boundary value problem for
the Stokes system
\begin{equation} \label{Stokes}
- \nu \Delta u + \nabla p = f, \quad \-\nabla\cdot u = g
\quad\mbox{in }{\cal G},
\end{equation}
i.e. a vector function $(u,p)\in W^{1,2}({\cal G})^3\times
L_2({\cal G})$ satisfying
\begin{eqnarray} \label{Stokes1}
&& b(u,v) - \int_{\cal G} p\, \nabla\cdot v\, dx = F(v)\ \mbox{
for all }v\in V, \\ \label{Stokes2} && -\nabla \cdot u =g \
\mbox{ in } {\cal G},\quad S_j u = h_j\ \mbox{ on }\Gamma_j,\
j=1,\ldots,N,
\end{eqnarray}
where $F$ is given by (\ref{a4}). For the proof of the following
theorem, we refer to \cite[Th.5.1]{mr-04b}.

\begin{Th} \label{Stokest1}
Let $g\in L_2({\cal G})$ and $h_j\in W^{1/2,2}(\Gamma_j)^{3-d_j}$
be such that there exists a vector function $v\in W^{1,2}({\cal
G})^3$, $S_j v=h_j$ on $\Gamma_j$, $j=1,\ldots,N$. In the case
when $d_j \in \{ 0,2\}$ for all $j$, we assume in addition that
\begin{equation} \label{1Stokest1}
\int_{\cal G} g\, dx + \sum_{j:\, d_j=0} \int_{\Gamma_j} h_j\cdot
n\, dx +
  \sum_{j:\, d_j=2} \int_{\Gamma_j} h_j\, dx =0.
\end{equation}
Furthermore, let the functional $F\in V^*$ satisfy the condition
\begin{equation} \label{2Stokest1}
F(v) = 0 \quad\mbox{for all } v\in L_V,
\end{equation}
where $L_V$ denotes the set of all $v\in V$ such that
$\varepsilon_{i,j}(v)=0$ for $i,j=1,2,3$. Then there exists a
solution $(u,p)\in W^{1,2}({\cal G})^3 \times L_2({\cal G})$ of
the problem {\em (\ref{Stokes1}), (\ref{Stokes2})}. Here $p$ is
uniquely determined if $d_j \in \{ 1,3\}$ for at least one $j$ and
unique up to constants if $d_j\in \{ 0,2\}$ for all $j$. The
vector function $u$ is unique up to elements from $L_V$.
\end{Th}

Note that $L_V$ contains only functions of the form $v=c+Ax$,
where $c$ is a constant vector and $A$ is a constant matrix,
$A=-A^t$ (rigid body motions). In particular, $\nabla\cdot v=0$
for $v\in L_V$. In most cases (e.g. if the Dirichlet condition is
given on at least one face $\Gamma_j$), the set $L_V$ contains
only the function $v=0$.

\subsection{Existence of solutions of the nonlinear problem}

Let the operator $Q$ be defined by
\[
Qu = (u\cdot \nabla)\, u.
\]
Obviously, $Q$ realizes a mapping $W^{1,2}({\cal G})\to V^*$.
Furthermore, there exist constants $c_1,c_2$ such that
\begin{eqnarray} \label{Q1}
&& \| Qu\|_{V^*} \le c\, \| u\|^2_{W^{1,2}({\cal G})^3}
\quad\mbox{for all }u\in W^{1,2}({\cal G}),\\ \label{Q2} && \|
Qu-Qv\|_{V^*} \le c\, \big( \| u\|_{W^{1,2}({\cal G})^3}+ \|
v\|_{W^{1,2}({\cal G})^3}\big) \,
  \| u-v\|_{W^{1,2}({\cal G})^3}   \quad\mbox{for all }u,v\in W^{1,2}({\cal G})^3.\qquad
\end{eqnarray}
Using the last two estimates together with Theorem \ref{Stokest1},
we can prove the following statement.

\begin{Th} \label{Stokest1a}
Let $g$ and $h_j$ be as in Theorem {\em \ref{Stokest1}}.
Furthermore, we suppose that $L_V=\{ 0\}$ and
\[
\| F\|_{V^*} + \| g\|_{L_2({\cal G})} + \sum_{j=1}^N \|
h_j\|_{W^{1/2,2}(\Gamma_j)^{3-d_j}}
\]
is sufficiently small. Then there exists a solution $(u,p)\in
W^{1,2}({\cal G})^3\times L_2({\cal G})$ of the problem {\em
(\ref{a1}), (\ref{a2})}. Here, $u$ is unique on the set of all
functions with norm less than a certain positive $\varepsilon$,
$p$ is unique if $d_j \in \{ 1,3\}$ for at least one $j$,
otherwise $p$ is unique up to a constant.
\end{Th}

P r o o f. Let $(u^{(0)},p^{(0)})\in W^{1,2}({\cal G})^3\times
L_2({\cal G})$ be the solution of the linear problem
(\ref{Stokes1}), (\ref{Stokes2}). By our assumptions on $F$, $g$
and $h_j$, we may assume that
\[
\| u^{(0)}\|_{W^{1,2}({\cal G})^3} \le \varepsilon_1,
\]
where $\varepsilon_1$ is a small positive number. Let $V_0$ denote
the set of all $v\in V$ such that $\nabla\cdot v=0$. We put
$w=u-u^{(0)}$ and $q=p-p^{(0)}$. Then $(u,p)$ is a solution of the
problem (\ref{a1}), (\ref{a2}) if and only if $(w,q)\in V_0\times
L_2({\cal G})$ and
\begin{equation} \label{1Stokest1a}
b(w,v)-\int_{\cal G} q\, \nabla\cdot v\, dx = -\int_{\cal G}
Q(w+u^{(0)})\cdot v\, dx \quad\mbox{for all }v\in V.
\end{equation}
By Theorem \ref{Stokest1}, there is a linear and continuous
mapping
\[
V^* \ni \Phi \to A\Phi =(w,q) \in V_0\times L_2({\cal G})
\]
defined by
\[
b(w,v)-\int_{\cal G} q\, \nabla\cdot v\, dx = \Phi(v)\ \mbox{ for
all }v\in V,\quad
  \int_{\cal G} q\, dx = 0 \ \mbox{if $d_j\in \{ 0,2\}$ for all $j$}.
\]
We write (\ref{1Stokest1a}) as
\[
(w,q) = T(w,q),\quad\mbox{where } T(w,q)=- AQ(w+u^{(0)}).
\]
Due to (\ref{Q2}), the operator $T$ is contractive on the set of
all $(w,q)\in V_0 \times L_2({\cal G})$ with norm $\le
\varepsilon_2$ if $\varepsilon_1$ and $\varepsilon_2$ are
sufficiently small. Hence there exist $w\in W^{1,2}({\cal G})^3$
and $q\in L_2({\cal G})$ satisfying (\ref{1Stokest1a}). The result
follows. \hfill $\Box$

\begin{Rem}
{\em If $d_j\in \{0,2\}$ for all $j$, then
\begin{equation} \label{trl}
\int_{\cal G} \sum_{j=1}^3 u_j\, \frac{\partial v}{\partial
x_j}\cdot v\, dx =0 \ \ \mbox{for all }v\in W^{1,2}({\cal G})^3,\
u\in V,\ \nabla\cdot u=0
\end{equation}
(see \cite[Le.IV.2.2]{Girault}). Thus, analogously to
\cite[Th.IV.2.3]{Girault} (see also \cite{Sol-73}), the problem
(\ref{a1}), (\ref{a2}) hast at least one solution for arbitrary
$F\in V^*$, $g=0$, $h_j\in W^{1/2,2}(\Gamma_j)^{3-d_j}$ satisfying
(\ref{1Stokest1}).}
\end{Rem}

\setcounter{equation}{0}
\section{Weighted Sobolev and H\"older spaces}

Here, we introduce weighted Sobolev and H\"older spaces in
polyhedral domains and prove imbeddings for these spaces which
will be used in the next section. We start with the case of a
polyhedral cone.

\subsection{Weighted Sobolev spaces in a cone}

Let ${\cal K} = \{ x\in {\Bbb R}^3:\, x/|x| \in \Omega\}$ be a
polyhedral cone in ${\Bbb R}^3$ whose boundary consists of plane
faces $\Gamma_j$ and edges $M_k$, $j,k=1,\ldots,N$. We denote by
$\rho(x)=|x|$ the distance of $x$ to the vertex of the cone, by
$r_k(x)$ the distance to the edge $M_k$, and by $r(x)$ the
distance to the set ${\cal S}=M_1\cup\cdots\cup M_N\cup\{ 0\}$.
Note that there exist positive constants $c_1$, $c_2$ such that
\begin{equation} \label{rx}
c_1\, r(x) \le \rho(x)\, \prod_{k=1}^N \frac{r_k(x)}{\rho(x)} \le
c_2\, r(x)\quad\mbox{for all }x\in {\cal K}.
\end{equation}
Let $l$ be a nonnegative integer, $\beta\in {\Bbb R}$,
$\delta=(\delta_1,\ldots,\delta_N)\in {\Bbb R}^N$, and
$1<s<\infty$. We define $V_{\beta,\delta}^{l,s}({\cal K})$ as the
closure of the set $C_0^\infty(\overline{\cal K}\backslash {\cal
S})$ with respect to the norm
\[
\| u\|_{V_{\beta,\delta}^{l,s}({\cal K})}= \Big( \int_{\cal K}
\sum_{|\alpha|\le l}
  \rho^{s(\beta-l+|\alpha|)} \, \prod_{k=1}^N \big(\frac{r_k}{\rho}\Big)^{s(\delta_k-l+|\alpha|)}
  \ |\partial_x^\alpha u|^s \, dx\Big)^{1/s}.
\]
The weighted Sobolev space $W_{\beta,\delta}^{l,s}({\cal K})$,
where $\delta_k>-2/s$ for $k=1,\ldots,N$, is defined as the
closure of the set $C_0^\infty(\overline{\cal K})$ with respect to
the norm
\[
\| u\|_{W_{\beta,\delta}^{l,s}({\cal K})} = \Big( \int_{\cal K}
  \sum_{|\alpha|\le l} \rho^{s(\beta-l+|\alpha|)}  \prod_{k=1}^N \big(\frac{r_k}{\rho}\big)^{s\delta_k}\, dx\Big)^{1/s}.
\]
If $\delta$ is a real number, then by
$V_{\beta,\delta}^{l,s}({\cal K})$ and
$W_{\beta,\delta}^{l,s}({\cal K})$, we mean the above introduced
spaces with $\delta_1=\cdots=\delta_N=\delta$. For the proof of
the following lemma we refer to \cite[Le.1]{mr-91}.

\begin{Le} \label{al2}
Let $1<s \le t<\infty$, $l-3/s \ge l'-3/t$, $\beta-l+3/s =
\beta'-l'+3/t$, and $\delta_k-l+3/s \le \delta'_k-l'+3/t$ for
$k=1,\ldots,N$. Then $V_{\beta,\delta}^{l,s}({\cal K})$ is
continuously imbedded into $V_{\beta',\delta'}^{l',t}({\cal K})$.
\end{Le}

In particular, we have $V_{\beta,\delta}^{l,s}({\cal K}) \subset
V_{\beta',\delta'}^{l',s} ({\cal K})$ for $l\ge l'$,
$\beta-l=\beta'-l'$ and $\delta_k-l\le \delta'_k-l'$,
$k=1,\ldots,N$. If in addition $\delta_k> -2/s$ and $\delta'_k
>-2/s$ for $k=1,\ldots,N$, then also
\[
W_{\beta,\delta}^{l,s}({\cal K}) \subset
W_{\beta',\delta'}^{l',s}({\cal K}).
\]
The spaces $V_{\beta,\delta}^{l,s}({\cal K})$ and
$W_{\beta,\delta}^{l,s}({\cal K})$ coincide if $\delta_k>l-2/s$
for all $k$ (see \cite{mr-04b}).

\begin{Le} \label{al3}
Let $1<s\le t<\infty$ and $l-3/s\ge \max(\delta_k,0)-3/t$. Then
$W_{\beta,\delta}^{l,s}({\cal K}) \subset
W_{\beta-l+3/s-3/t,0}^{0,t}({\cal K})$ and
\begin{equation} \label{1al3}
\| \rho^{\beta-l+3/s-3/t}u\|_{L_t({\cal K})} \le c\, \|
u\|_{W_{\beta,\delta}^{l,s}({\cal K})}
\end{equation}
for all $u\in W_{\beta,\delta}^{l,s}({\cal K})$ with a constant
$c$ independent of $u$. Furthermore, for arbitrary $u \in
W_{\beta,\delta}^{l,s}({\cal K})$, $l-3/s >\max(\delta_k,0)$, the
following inequality is valid.
\begin{equation} \label{2al3}
\| \rho^{\beta-l+3/s}u\|_{L_\infty({\cal K})} \le c\, \|
u\|_{W_{\beta,\delta}^{l,s}({\cal K})}
\end{equation}
\end{Le}

P r o o f. Let $u \in W_{\beta,\delta}^{l,s}({\cal K})$, and let
$\zeta_k$ be infinitely differentiable functions with support in
$\{ x:\, 2^{k-1}< |x|<2^{k+1}\}$ such that
\[
\big| \partial_x^\alpha \zeta_k(x)\big| < c\, 2^{-k|\alpha|} \
\mbox{ for }k\le l,\quad
  \sum_{k=-\infty}^{+\infty} \zeta_k =1.
\]
We define $v(x)=u(2^k x)$ and $\eta_k(x)=\zeta_k(2^k x)$. Then
$\eta_k(x)$ vanishes for $|x|<1/2$ and $|x|>2$. Therefore for
$l'=l-\max(\delta_k,0)$, we have
\[
\| \eta_k v\|_{W^{l',s}({\cal K})} \le c\, \| \eta_k
v\|_{W_{\beta,\delta}^{l,s}({\cal K})}
\]
(see \cite[Th.3]{r-92}). This inequality together with the
continuity of the imbedding $W^{l',s} \subset L_t$ implies
\[
\| \eta_k v\|_{L_t({\cal K})} \le c\, \| \eta_k
v\|_{W_{\beta,\delta}^{l,s}({\cal K})},
\]
where $c$ is independent of $u$ and $k$. Using the equalities
\[
\| \eta_k v\|_{L_t({\cal K})} = 2^{-3k/t} \, \| \zeta_k
u\|_{L_t({\cal K})}\quad\mbox{and}\quad
  \| \eta_k v\|_{W_{\beta,\delta}^{l,s}({\cal K})} = 2^{-k(\beta-l)-3k/s}
  \, \| \zeta_k u\|_{W_{\beta,\delta}^{l,s}({\cal K})}\, ,
\]
we obtain
\[
\| \rho^{\beta-l+3/s-3/t}\zeta_k u\|_{L_t({\cal K})} \le c\, \|
\zeta_k u\|_{W_{\beta,\delta}^{l,s}({\cal K})}\, .
\]
Consequently,
\begin{eqnarray*}
\| \rho^{\beta-l+3/s-3/t}u\|_{L_t({\cal K})} & \le & c\, \Big(
\sum_{k=-\infty}^{+\infty} \| \rho^{\beta-l+3/s-3/t}
  \zeta_k u\|^t_{L_t({\cal K})}\Big)^{1/t} \le c\, \Big( \sum_{k=-\infty}^{+\infty}
  \| \zeta_k u \|^t_{W_{\beta,\delta}^{l,s}({\cal K})}\Big)^{1/t} \\
& \le & c\, \Big( \sum_{k=-\infty}^{+\infty}  \| \zeta_k
u\|^s_{W_{\beta,\delta}^{l,s}({\cal K})}\Big)^{1/s}
  \le c\, \| \zeta_k u\|_{W_{\beta,\delta}^{l,s}({\cal K})} \, .
\end{eqnarray*}
This proves (\ref{1al3}). The proof of (\ref{2al3}) proceeds
analogously. \hfill $\Box$

\begin{Le} \label{al4}
Let $u\in W_{\beta,\delta}^{l,s}({\cal K})$, and let $j$ be an
integer, $j\ge 1$. Then there exist functions $v\in
V_{\beta,\delta}^{l,s}({\cal K})$ and $w\in
W_{\beta+j,\delta+j}^{l+j,s}({\cal K})$ such that $u=v+w$ and
\[
\| v\|_{V_{\beta,\delta}^{l,s}({\cal K})} + \|
w\|_{W_{\beta+j,\delta+j}^{l+j,s}({\cal K})}
  \le c\, \| u\|_{W_{\beta,\delta}^{l,s}({\cal K})}\, .
\]
\end{Le}

P r o o f. Let $\zeta_k$ be the same functions as in the proof of
Lemma \ref{al3}, and let $u$ be an arbitrary function from
$W_{\beta,\delta}^{l,s}({\cal K})$. Obviously, the function
$\tilde{u}_k$ defined by $\tilde{u}_k(x) = \zeta_k(2^k x)\, u(2^k
x)$ belongs also to $W_{\beta,\delta}^{l,s}({\cal K})$ and
vanishes for $|x|<1/2$ and $|x|>2$. Consequently by
\cite[Le.1.3]{mr-88} (for integer $\beta+2/s$ see
\cite[Th.5]{r-92}), there exist functions $\tilde{v}_k \in
V_{\beta,\delta}^{l,s}({\cal K})$ and $\tilde{w}_k \in
W_{\beta+j,\delta+j}^{l+j,s} ({\cal K})$ with supports in $\{x:\,
1/4<|x|<4\}$ such that $\tilde{u}_k=\tilde{v}_k + \tilde{w}_k$ and
\[
\| \tilde{v}_k\|_{V_{\beta,\delta}^{l,s}({\cal K})} + \|
\tilde{w}_k\|_{W_{\beta+j,\delta+j}^{l+j,s}({\cal K})}
  \le c\, \| \tilde{u}_k \|_{W_{\beta,\delta}^{l,s}({\cal K})},
\]
where $c$ is independent of $u$ and $k$. Let
$v_k(x)=\tilde{v}_k(2^{-k}x)$ and $w_k(x)=\tilde{w}_k(2^{-k}x)$.
Then the supports of $v_k$ and $w_k$ are contained in $\{x:\,
2^{k-2}<|x|<2^{k+2}\}$, and we have $\zeta_k u= v_k+w_k$ for all
$k$. Moreover,
\begin{eqnarray*}
\| v_k\|_{V_{\beta,\delta}^{l,s}({\cal K})} + \|
w_k\|_{W_{\beta+j,\delta+j}^{l+j,s}({\cal K})} & = &
2^{k(\beta-l+3/s)} \Big( \|
\tilde{v}_k\|_{V_{\beta,\delta}^{l,s}({\cal K})}
  + \| \tilde{w}_k\|_{W_{\beta+j,\delta+j}^{l+j,s}({\cal K})}\Big)\\
& \le  & c\, 2^{k(\beta-l+3/s)}  \, \| \tilde{u}_k
\|_{W_{\beta,\delta}^{l,s}({\cal K})}
  = c\, \| \zeta_k u \|_{W_{\beta,\delta}^{l,s}({\cal K})}\, .
\end{eqnarray*}
Let $v=\sum_{k=-\infty}^{+\infty} v_k$ and
$w=\sum_{k=-\infty}^{+\infty} w_k$. Then $u=v+w$. Since $v_k$ and
$v_m$  have disjoint supports for $|k-m|\ge 4$, we have $v\in
V_{\beta,\delta}^{l,s}({\cal K})$ and
\[
\| v\|^s_{V_{\beta,\delta}^{l,s}({\cal K})} \le 7^{s-1} \, \sum_k
\| v_k \|^s_{V_{\beta,\delta}^{l,s}({\cal K})}
  \le c\, \sum_k \| \zeta_k u \|^s_{W_{\beta,\delta}^{l,s}({\cal K})} \le c'\,
  \| u \|^s_{W_{\beta,\delta}^{l,s}({\cal K})}.
\]
Analogously, the norm of $w$ in
$W_{\beta+j,\delta+j}^{l+j,s}({\cal K})$ can be estimated by the
norm of $u$ in $W_{\beta,\delta}^{l,s}({\cal K})$. The lemma is
proved. \hfill $\Box$

\subsection{Weighted H\"older spaces in a polyhedral cone}

Let $l$ be a nonnegative integer, $\beta\in {\Bbb R}$,
$\delta=(\delta_1,\ldots,\delta_N)\in {\Bbb R}^N$, and $\sigma\in
(0,1)$. We define the weighted H\"older space ${\cal
N}_{\beta,\delta}^{l,\sigma}({\cal K})$ as the set of all $l$
times continuously differentiable functions on $\bar{\cal
K}\backslash {\cal S}$ with finite norm
\begin{eqnarray} \label{LambdaK}
\| u\|_{{\cal N}_{\beta,\delta}^{l,\sigma}({\cal K})} & \!\! =
&\!\!
  \sum_{|\alpha|\le l} \sup_{x\in{\cal K}} |x|^{\beta-l-\sigma+|\alpha|}
  \prod_{k=1}^N \Big( \frac{r_k(x)}{|x|}\Big)^{\delta_k-l-\sigma+|\alpha|}
  \  \big| \partial_x^\alpha u(x)\big| \nonumber\\
&& + \sum_{|\alpha|=l} \sup_{|x-y|<r(x)/2} |x|^\beta\,
\prod_{k=1}^N \Big( \frac{r_k(x)}{|x|}\Big)^{\delta_k}
  \frac{\big|\partial_x^\alpha  u(x)-\partial_y^\alpha u(y)\big|}{|x-y|^\sigma} \, .
\end{eqnarray}
Furthermore, the space $C_{\beta,\delta}^{l,\sigma}({\cal K})$ is
defined for nonnnegative $\delta_k$, $k=1,\ldots,N$, as the set of
all $l$ times continuously differentiable functions on $\bar{\cal
K}\backslash {\cal S}$ with finite norm
\begin{eqnarray} \label{CK}
\| u\|_{C_{\beta,\delta}^{l,\sigma}({\cal K})} & \!\! = &\!\!
  \sum_{|\alpha|\le l} \sup_{x\in{\cal K}} |x|^{\beta-l-\sigma+|\alpha|}
  \prod_{k=1}^N \Big( \frac{r_k(x)}{|x|}\Big)^{\max(0,\delta_k-l-\sigma+|\alpha|)}
  \  \big| \partial_x^\alpha u(x)\big| \nonumber\\
&& + \sum_{k:\, \sigma_k\le l} \ \sum_{|\alpha|=l-\sigma_k}
\sup_{\substack{x,y\in {\cal K}_k \\ |x-y|<|x|/2}}
  |x|^{\beta-\delta_k} \frac{\big|\partial_x^\alpha  u(x)-\partial_y^\alpha u(y)\big|}
  {|x-y|^{\sigma+\sigma_k-\delta_k}} \nonumber \\
&& + \sum_{|\alpha|=l} \sup_{\substack{x,y\in {\cal K}\\
|x-y|<r(x)/2}} |x|^\beta\, \prod_{k=1}^N \Big(
  \frac{r_k(x)}{|x|}\Big)^{\delta_k}\frac{\big|\partial_x^\alpha  u(x)-\partial_y^\alpha u(y)\big|}{|x-y|^\sigma} \, ,
\end{eqnarray}
where ${\cal K}_k = \{ x\in {\cal K}:\, r_k(x)<3r(x)/2\}$,
$\sigma_k=[\delta_k-\sigma]+1$, $[s]$ denotes the greatest integer
less or equal to $s$. The trace spaces for ${\cal
N}_{\beta,\delta}^{l,\sigma}({\cal K})$ and
$C_{\beta,\delta}^{l,\sigma}({\cal K})$ on $\Gamma_j$ are denoted
by ${\cal N}_{\beta,\delta}^{l,\sigma}(\Gamma_j)$ and
$C_{\beta,\delta}^{l,\sigma}(\Gamma_j)$, respectively.

Obviously, ${\cal N}_{\beta,\delta}^{l,\sigma}({\cal K})$ is a
subset of $C_{\beta,\delta}^{l,\sigma}({\cal K})$. If $\delta_k\ge
l+\sigma$ for $k=1,\ldots,N$, then both spaces coincide.
Furthermore, the following imbedding holds (and is continuous, see
\cite{mr-05}).
\[
{\cal N}_{\beta,\delta}^{l,\sigma}({\cal K}) \subset {\cal
N}_{\beta',\delta'}^{l',\sigma'}({\cal K})\
  \mbox{ if  } l+\sigma \ge l'+\sigma',\ \beta-l-\sigma=\beta'-l'-\sigma', \
  \delta_k-l-\sigma\le \delta'_k-l'-\sigma' .
\]
If in addition $\delta_k$ and $\delta'_k$ are nonnegative, then
$C_{\beta,\delta}^{l,\sigma}({\cal K})$ is continuously imbedded
into $C_{\beta',\delta'}^{l',\sigma'}({\cal K})$. Next, we prove a
relation between the spaces $V_{\beta,\delta}^{l,s}$ and ${\cal
N}_{\beta,\delta}^{l,\sigma}$.

\begin{Le} \label{al5}
Suppose that $l-3/s > l'+\sigma$, $\beta-l+3/s=\beta'-l'-\sigma$
and $\delta_k-l+3/s\le \delta'_k-l'-\sigma$ for $k=1,\ldots,N$.
Then $V_{\beta,\delta}^{l,s}({\cal K})$ is continuously imbedded
into ${\cal N}_{\beta',\delta'}^{l',\sigma }({\cal K})$.
\end{Le}

P r o o f. It suffices to prove the lemma for $\delta_k-l+3/s =
\delta'_k-l'-\sigma$, $k=1,\ldots,N$. Let $u \in
V_{\beta,\delta}^{l,s}({\cal K})$. For an arbitrary point $x\in
{\cal K}$, we denote by $B_x$ the set $\{x'\in {\cal K}:\,
|x-x'|<r(x)/2\}.$ Note that
\begin{equation} \label{1al5}
|x|/2 \le |x'|< 3|x|/2,\quad r_k(x)/2 \le r_k(x') < 3r_k(x)/2,
\quad
  r(x)/2\le r(x')\le 3r(x)/2 \quad \mbox{for } x'\in B_x.
\end{equation}
First, let $r(x)=1$. From the continuity of the imbedding
$W^{l,s}(B_x)  \subset C^{l',\sigma}(B_x)$ it follows that there
exists a constant $c$ independent of $u$ and $x$ such that
\begin{eqnarray*}
&& \big|(\partial^\alpha u)(x)\big| \le c\, \| u\|_{W^{l,s}(B_x)}\quad\mbox{for }|\alpha|\le l',\\
&& \frac{|(\partial^\alpha u)(x)-(\partial^\alpha
u)(x')|}{|x-x'|^\sigma} \le c\, \| u\|_{W^{l,s}(B_x)}
  \quad\mbox{for }|\alpha|= l',\ x'\in B_x\, .
\end{eqnarray*}
Due to (\ref{rx}) and (\ref{1al5}), this implies
\[
|x|^{\beta} \, \prod_k \big( \frac{r_k(x)}{|x|}\big)^{\delta_k}
  \big|(\partial^\alpha u)(x)\big| \le c\, \sum_{|\gamma|\le l} \Big\| r^{|\gamma|-l}\, \rho^{\beta} \,
  \prod_k \big( \frac{r_k}{\rho}\big)^{\delta_k}\, \partial^\gamma u\Big\|_{L_s(B_x)}
  \le c\, \| u\|_{V_{\beta,\delta}^{l,s}({\cal K})} \quad\mbox{for }|\alpha|\le l'
\]
and analogously
\[
|x|^{\beta} \, \prod_k \big( \frac{r_k(x)}{|x|}\big)^{\delta_k} \,
  \frac{|(\partial^\alpha u)(x)-(\partial^\alpha u)(x')|}{|x-x'|^\sigma}
  \le c\, \| u\|_{V_{\beta,\delta}^{l,s}({\cal K})} \quad\mbox{for }|\alpha|=l',\ x'\in B_x.
\]
Now let $x$ be an arbitrary point in ${\cal K}$ and $x'\in B_x$.
We put $y= x/r(x)$, $y'=x'/r(x)$. Then $r(y)=1$ and $y'\in B_y$.
Consequently, the function $v(\xi)=u\big(r(x)\, \xi\big)$
satisfies the inequalities
\[
|y|^{\beta} \, \prod_k \big( \frac{r_k(y)}{|y|}\big)^{\delta_k}
  \big| (\partial^\alpha v)(y)\big| \le c\, \| v\|_{V_{\beta,\delta}^{l,s}({\cal K})}
  \le c\, r(x)^{l-\beta-3/s}\, \| u \|_{V_{\beta,\delta}^{l,s}({\cal K})}
\]
for $|\alpha|\le l'$ and
\[
|y|^{\beta} \, \prod_k \big( \frac{r_k(y)}{|y|}\big)^{\delta_k} \,
  \frac{|( \partial^\alpha v)(y)-(\partial^\alpha v)(y')|}{|y-y'|^\sigma}
  \le c\, r(x)^{l-\beta-3/s}\, \| u\|_{V_{\beta,\delta}^{l,s}({\cal K})}.
\]
for $|\alpha|=l'$. Here, the constant $c$ is independent of $u$,
$x$ and $x'$. Using (\ref{rx}), we obtain the inequalities
\begin{eqnarray*}
&& |x|^{\beta-l+|\alpha|+3/s} \, \prod_k \big(
\frac{r_k(x)}{|x|}\big)^{\delta_k-l+|\alpha|+3/s}
  \big|(\partial^\alpha u)(x)\big| \le c\, \| u \|_{V_{\beta,\delta}^{l,s}({\cal K})}
  \quad\mbox{for }|\alpha|\le l', \\
&& |x|^{\beta'} \, \prod_k \big(
\frac{r_k(x)}{|x|}\big)^{\delta_k-l+l'+\sigma+3/s} \,
  \frac{|(\partial^\alpha u)(x)-(\partial^\alpha u)(x')|}{|x-x'|^\sigma}
  \le c\, \| u\|_{V_{\beta,\delta}^{l,s}({\cal K})} \quad\mbox{for }|\alpha|=l'.
\end{eqnarray*}
The result follows. \hfill $\Box$

\begin{Co} \label{ac2}
Let $u\in W_{\beta,\delta}^{l,s}({\cal K})$, $l>3/s$. Then
\begin{equation} \label{1ac2}
\rho^{\beta-l+3/s}\, \prod_k \big( \frac{r_k}\rho
\big)^{\sigma_k}\, u \in L_\infty({\cal K}),
\end{equation}
where $\sigma_k=0$ for $\delta_k<l-3/s$,
$\sigma_k=1/s+\varepsilon$ for $l-3/s\le \delta_k\le l-2/s$, and
$\sigma_k=\delta_k-l+3/s$ for $\delta_k>l-2/s$ ($\varepsilon$ is
an arbitrarily small positive number).
\end{Co}

P r o o f. Let $\psi_k$ be a smooth function on the unit sphere
$S^2$ such that $\psi_k=\delta_{j,k}$ in a neighborhood of the
points $S^2\cap M_j$ for $j=1,\ldots,m$. We extend $\psi_k$ to
${\Bbb R}^3\backslash \{ 0\}$ by $\psi_k(x)=\psi_k(x/|x|)$. Then
$\psi_k u \in W_{\beta,\delta_k}^{l,s}({\cal K})$. Obviously, it
suffices to prove (\ref{1ac2}) for the function $\psi_k u$. If
$\delta_k<l-3/s$, then by Lemma \ref{al3},
$\rho^{\beta-l+3/s}\psi_k u\in L_\infty({\cal K})$. If
$\delta_k>l-2/s$, then $W_{\beta,\delta_k}^{l,s}({\cal
K})=V_{\beta,\delta_k}^{l,s}({\cal K})$. By Lemma \ref{al5}, the
last space is imbedded into ${\cal
N}^{0,\sigma}_{\beta-l+\sigma+3/s,\delta_k-l+\sigma+3/s}({\cal
K})$ for arbitrary $\sigma<l-3/s$. Therefore in particular,
\[
\rho^{\beta-l+3/s} \prod_k \big( \frac{r_k}\rho
\big)^{\delta_k-l+3/s} \psi_k u \in L_\infty({\cal K}).
\]
For $l-3/s\le \delta_k<l-2/s$, the assertion follows from the
imbedding $W_{\beta,\delta_k}^{l,s}({\cal K})
\subset W_{\beta,l-2/s+\varepsilon}^{l,s}({\cal K})$. \hfill $\Box$ \\

\subsection{Weighted Sobolev and H\"older spaces in a bounded polyhedral domain}

Let ${\cal G}$ be a domain of polyhedral type (see Section 1.1)
with faces $\Gamma_1,\ldots,\Gamma_N$, edges $M_1,\ldots,M_m$ and
vertices $x^{(1)},\ldots,x^{(d)}$. We denote the distance of $x$
to the edge $M_k$ by $r_k(x)$, the distance to the vertex
$x^{(j)}$ by $\rho_j(x)$, the distance to ${\cal S}$ (the set of
all edge points and vertices) by $r(x)$, and the distance to the
set $X=\{ x^{(1)},\ldots,x^{(d)}\}$ by $\rho(x)$. For arbitrary
integer $l\ge 0$, real $s>1$ and real tuples
$\beta=(\beta_1,\ldots,\beta_d)$,
$\delta=(\delta_1,\ldots,\delta_m)$, we define
$V_{\beta,\delta}^{l,s}({\cal G})$ and
$W_{\beta,\delta}^{l,s}({\cal G})$ as the weighted Sobolev spaces
with the norms
\begin{eqnarray*}
&& \| u\|_{V_{\beta,\delta}^{l,s}({\cal G})}= \Big( \int_{\cal K}
\sum_{|\alpha|\le l}\
  \prod_{j=1}^d \rho_j^{s(\beta_j-l+|\alpha|)} \ \prod_{k=1}^m \big(\frac{r_k}{\rho}
  \Big)^{s(\delta_k-l+|\alpha|)} \ |\partial_x^\alpha u|^s \, dx\Big)^{1/s}, \\
&& \| u\|_{W_{\beta,\delta}^{l,s}({\cal K})} = \Big( \int_{\cal K}
  \sum_{|\alpha|\le l} \, \prod_{j=1}^d \rho_j^{s(\beta_j-l+|\alpha|)} \
  \prod_{k=1}^m \big(\frac{r_k}{\rho}\big)^{s\delta_k} \ |\partial_x^\alpha u|^s\, dx\Big)^{1/s},
\end{eqnarray*}
respectively. In the case of the space
$W_{\beta,\delta}^{l,s}({\cal G})$, we suppose that
$\delta_k>-2/s$ for $k=1,\ldots,m$. The corresponding trace spaces
on the faces $\Gamma_j$ are denoted by
$V_{\beta,\delta}^{l-1/s,s}(\Gamma_j)$ and
$W_{\beta,\delta}^{l-1/s,s}(\Gamma_j)$, respectively. Furthermore,
let $V_{\beta,\delta}^{-1,s}({\cal G})$ denote the dual space of
$V_{-\beta,-\delta}^{1,s'}({\cal G})$, $s'=s/(s-1)$, with respect
to the $L_2$ scalar product.

\begin{Le} \label{al7}
Let $1<t< s<\infty$, $\beta_j+3/s < \beta'_j+3/t$ for
$j=1,\ldots,d$, and $\delta_k+2/s <\delta'_k+2/t$ for
$k=1,\ldots,N$. Then $V_{\beta,\delta}^{l,s}({\cal G}) \subset
V_{\beta',\delta'}^{l,t}({\cal G})$ and
$W_{\beta,\delta}^{l,s}({\cal G}) \subset
W_{\beta',\delta'}^{l,t}({\cal G})$. These imbeddings are
continuous.
\end{Le}

P r o o f. Let $q=st/(s-t)$. By H\"older's inequality,
\[
\Big\| \prod_j \rho_j^{\beta'_j-l+|\alpha|} \prod_k \big(
\frac{r_k}\rho \big)^{\delta'_k} \, \partial_x^\alpha u
  \Big\|_{L_t({\cal G})} \le c\, \Big\| \prod_j \rho_j^{\beta_j-l+|\alpha|} \prod_k \big( \frac{r_k}\rho
  \big)^{\delta_k} \, \partial_x^\alpha u   \Big\|_{L_t({\cal G})}\, ,
\]
where
\[
c= \Big\| \prod_j \rho_j^{\beta'_j-\beta_j} \prod_k \big(
\frac{r_k}\rho \big)^{\delta'_k-\delta_k}
  \Big\|_{L_q({\cal G})} < \infty
\]
if $\beta'_j-\beta_j>-3/q$ and $\delta'_k-\delta_k>-2/q$. This
proves the imbedding $W_{\beta,\delta}^{l,s}({\cal G}) \subset
W_{\beta',\delta'}^{l,t}({\cal G})$. Analogously, the
imbedding $V_{\beta,\delta}^{l,s}({\cal G}) \subset V_{\beta',\delta'}^{l,t}({\cal G})$ holds. \hfill $\Box$ \\

The following result can be directly deduced from Lemma \ref{al2}.

\begin{Le} \label{al8}
Let $1<s \le t<\infty$, $l-3/s \ge l'-3/t$, $\beta_j-l+3/s \le
\beta'_j-l'+3/t$ for $j=1,\ldots,d$, and $\delta_k-l+3/s \le
\delta'_k-l'+3/t$ for $k=1,\ldots,m$. Then
$V_{\beta,\delta}^{l,s}({\cal G})$ is continuously imbedded into
$V_{\beta',\delta'}^{l',t}({\cal G})$.
\end{Le}

\begin{Co} \label{ac1}
Let $1<s \le t<\infty$, $3/s \le 1+3/t$, $\beta_j+3/s \le
\beta'_j+1+3/t$ for $j=1,\ldots,d$, and $\delta_k+3/s \le
\delta'_k+1+3/t$ for $k=1,\ldots,N$. Then
$V_{\beta,\delta}^{0,s}({\cal G})$ is continuously imbedded into
$V_{\beta',\delta'}^{-1,t}({\cal G})$.
\end{Co}

P r o o f. According to Lemma \ref{al8}, we have
$V_{-\beta',-\delta'}^{1,t'}({\cal G}) \subset
V_{-\beta,-\delta}^{0,s'}({\cal G})$, where $s'=s/(s-1)$, $t'=t/(t-1)$. The result follows. \hfill $\Box$\\

Let us further note that (as in the case of a cone)
\[
W_{\beta,\delta}^{l,s}({\cal G}) \subset
W_{\beta',\delta'}^{l',s}({\cal G})\quad \mbox{if }
  l\ge l',\ \beta_j-l\le \beta'_j-l',\ \delta_k-l\le \delta'_k-l',\ \delta_k>-2/s,\ \delta'_k>-2/s
\]
for $j=1,\ldots,d$, $k=1,\ldots,m$. If $\delta_k>l-2/s$ for
$k=1,\ldots,m$, then $V_{\beta,\delta}^{l,s}({\cal G})
= W_{\beta,\delta}^{l,s}({\cal G})$. \\

We introduce the following weighted H\"older spaces in the domain
${\cal G}$. The space ${\cal N}_{\beta,\delta}^{l,\sigma}({\cal
G})$ is defined as the set all $l$ times continuously
differentiable functions on $\bar{\cal G}\backslash {\cal S}$
with finite norm
\begin{eqnarray*}
\| u\|_{{\cal N}_{\beta,\delta}^{l,\sigma}({\cal G})} & \!\! =
&\!\!
  \sum_{|\alpha|\le l}\ \sup_{x\in{\cal G}} \ \prod_{j=1}^d \rho_j(x)^{\beta_j-l-\sigma+|\alpha|}
  \prod_{k=1}^{m} \Big( \frac{r_k(x)}{\rho(x)}\Big)^{\delta_k-l-\sigma+|\alpha|}
  \  \big| \partial_x^\alpha u(x)\big| \\
&& + \sum_{|\alpha|=l} \sup_{|x-y|<r(x)/2} \prod_{j=1}^d
\rho_j(x)^{\beta_j}\,
  \prod_{k=1}^{m} \Big( \frac{r_k(x)}{\rho(x)}\Big)^{\delta_k}
  \frac{\big|\partial_x^\alpha  u(x)-\partial_y^\alpha u(y)\big|}{|x-y|^\sigma} \, .
\end{eqnarray*}
Suppose that $\delta_k\ge 0$ for $k=1,\ldots,m$. Then
$C_{\beta,\delta}^{l,\sigma}({\cal G})$, is defined as the set of
all $l$ times continuously differentiable functions on $\bar{\cal
G}\backslash {\cal S}$  with finite norm
\begin{eqnarray*}
\| u\|_{C_{\beta,\delta}^{l,\sigma}({\cal G})} & \!\! = &\!\!
  \sum_{|\alpha|\le l}\ \sup_{x\in{\cal G}} \ \prod_{j=1}^d \rho_j(x)^{\beta_j-l-\sigma+|\alpha|}
  \prod_{k=1}^{m} \Big( \frac{r_k(x)}{\rho(x)}\Big)^{\max(0,\delta_k-l-\sigma+|\alpha|)}
  \  \big| \partial_x^\alpha u(x)\big| \\
&& + \sum_{j=1}^d  \ \sum_{k:\ \sigma_k\le l}
\sum_{|\alpha|=l-\sigma_k}
  \sup_{\substack{x,y\in {\cal G}_{j,k} \\ |x-y|<\rho(x)/2}}
  \rho_j(x)^{\beta_j-\delta_k} \frac{\big|\partial_x^\alpha  u(x)-\partial_y^\alpha u(y)\big|}
  {|x-y|^{\sigma+\sigma_k-\delta_k}} \\
&& + \sum_{|\alpha|=l} \sup_{|x-y|<r(x)/2} \prod_{j=1}^d
\rho_j(x)^{\beta_j}\,
  \prod_{k=1}^{m} \Big( \frac{r_k(x)}{\rho(x)}\Big)^{\delta_k}
  \frac{\big|\partial_x^\alpha  u(x)-\partial_y^\alpha u(y)\big|}{|x-y|^\sigma} \, ,
\end{eqnarray*}
where ${\cal G}_{j,k}=\{ x\in {\cal G}:\, \rho_j(x) < 3\rho(x)/2,\
r_k(x)<3r(x)/2\}$ and $\sigma_k=[\delta_k-\sigma]+1$. The trace
space on $\Gamma_j$ for $C_{\beta,\delta}^{l,\sigma}({\cal G})$ is
denoted by $C_{\beta,\delta}^{l,\sigma}(\Gamma_j)$.

Analogously to the case when the domain is a cone, we have
\[
{\cal N}_{\beta,\delta}^{l,\sigma}({\cal K}) \subset {\cal
N}_{\beta',\delta'}^{l',\sigma'}({\cal K})\
  \mbox{ if  } l+\sigma \ge l'+\sigma',\ \beta_j-l-\sigma\le \beta'_j-l'-\sigma', \
  \delta_k-l-\sigma\le \delta'_k-l'-\sigma'
\]
for $j=1,\ldots,d$, $k=1,\ldots,m$. If in addition $\delta_k$ and
$\delta'_k$ are nonnegative, then
$C_{\beta,\delta}^{l,\sigma}({\cal G}) \subset
C_{\beta',\delta'}^{l',\sigma'}({\cal G})$. Furthermore, it
follows from Lemma \ref{al5} that
\[
V_{\beta,\delta}^{l,s}({\cal G}) \subset {\cal
N}_{\beta',\delta'}^{l',\sigma }({\cal G}) \quad \mbox{if }
  l-3/s > l'+\sigma,\ \beta_j-l+3/s\le \beta'_j-l'-\sigma, \ \delta_k-l+3/s\le \delta'_k-l'-\sigma
\]
for $j=1,\ldots,d$, $k=1,\ldots,m$.

We introduce the following notation. If $\beta\in {\Bbb R}^d, \
\delta\in {\Bbb R}^{m}$ and $s,t\in {\Bbb R}$, then by ${\cal
N}_{\beta+s,\delta+t}^{l,\sigma}({\cal G})$ and
$C_{\beta+s,\delta+t}^{l,\sigma}({\cal G})$, we mean the spaces
${\cal N}_{\beta',\delta'}^{l,\sigma}({\cal G})$ and
$C_{\beta',\delta'}^{l,\sigma}({\cal G})$ with
$\beta'=(\beta_1+s,\ldots,\beta_d+s)$,
$\delta'=(\delta_1+t,\ldots,\delta_{m}+t)$. Analogous notation
will be used for the weighted Sobolev spaces
$V_{\beta,\delta}^{l,s}$ and $W_{\beta,\delta}^{l,s}$.

The next lemma follows immediately from the definition of the
space ${\cal N}_{\beta,\delta}^{l,\sigma}({\cal G})$.

\begin{Le} \label{al6}
If $f\in {\cal N}_{\beta,\delta}^{l,\sigma}({\cal G})$ and $g \in
{\cal N}_{\beta',\delta'}^{l,\sigma}({\cal G})$, then $fg \in
{\cal
N}_{\beta+\beta'-l-\sigma,\delta+\delta'-l-\sigma}^{l,\sigma}({\cal
G})$.
\end{Le}

Finally, we define $C_{\beta,\delta}^{-1,\sigma}({\cal G})$ as the
space of all distributions of the form
\begin{equation} \label{Cminus}
f= f_0 + \sum_{j=1}^3 \partial_{x_j}f_j,\quad \mbox{where } \
  f_0 \in C_{\beta+1,\delta+1}^{0,\sigma}({\cal G})\ \mbox{ and } \
  f_j \in C_{\beta,\delta}^{0,\sigma}({\cal G}),\ j=1,2,3.
\end{equation}
Note that every $f \in C^{-1,\sigma}({\cal G})$, i.e. every
distribution of the form
\begin{equation} \label{Cmin}
f= f_0 + \sum_{j=1}^3 \partial_{x_j}f_j,\quad \mbox{where } \  f_j
\in C^{0,\sigma}({\cal G}),\ j=0,1,2,3,
\end{equation}
belongs to $C_{0,0}^{-1,\sigma}({\cal G})$. Indeed, let $\chi_k$
be infinitely differentiable cut-off functions equal to one near
$x^{(k)}$ and to zero near the vertices $x^{(l)}$, $l\not=k$. Then
the distribution (\ref{Cmin}) can be written as
\[
f = F_0 + \sum_{j=1}^3 \partial_{x_j}F_j,\ \mbox{ where } F_0(x)=
f_0(x)-\sum_{k=1}^d \sum_{j=1}^3
  f_j(x^{(k)})\, \partial_{x_j}\chi_k(x), \ \ F_j(x)=f_j(x)-\sum_{k=1}^d \chi_k(x)\, f_j(x^{(k)}),
\]
$j=1,2,3$. Here, $F_0 \in C^{0,\sigma}({\cal G}) \subset
C_{1,1}^{0,\sigma}({\cal G})$, and from $F_j \in
C^{0,\sigma}({\cal G})$, $F_j(x^{(k)})=0$ it follows that $F_j \in
C_{0,0}^{0,\sigma}({\cal G})$ for $j=1,2,3$.

\setcounter{equation}{0}
\section{Regularity results for weak solutions}

In this section, we establish regularity results for weak
solutions in weighted Sobolev and H\"older spaces. The regularity
assertions are formulated in terms of eigenvalues of operator
pencils generated by the boundary value problem at the edge points
and vertices of the domain.

\subsection{Operator pencils generated by the boundary value problem}

We introduce the operator pencils generated by the problem
(\ref{NS1}), (\ref{NS2}) for the edge points and vertices of the
domain ${\cal G}$.

1) Let $\xi$ be a point on an edge $M_k$, and let
$\Gamma_{k_+},\Gamma_{k_-}$ be the faces of ${\cal G}$ adjacent to
$\xi$. Then by ${\cal D}_\xi$ we denote the dihedron which is
bounded by the half-planes $\Gamma_{k_\pm}^\circ$ tangent to
$\Gamma_{k_\pm}$ at $\xi$ and the edge
$M_\xi^\circ=\bar{\Gamma}_{k_+}^\circ \cap
\bar{\Gamma}_{k_-}^\circ$. The angle between the half-planes
$\Gamma_{k_\pm}^\circ$ is denoted by $\theta_\xi$. Furthermore,
let $r,\varphi$ be polar coordinates in the plane perpendicular to
$M_\xi^\circ$ such that
\[
\Gamma_{k_\pm}^\circ = \{ x\in {\Bbb R}^3: \ r>0,\ \varphi=\pm
\theta_\xi/2\}.
\]
Then we define the operator $A_\xi(\lambda)$ as follows:
\[
A_\xi(\lambda)\, \big( U(\varphi) , P(\varphi)\big) = \big(
  r^{2-\lambda}(-\Delta u+ \nabla p)\, ,\, -r^{1-\lambda}\nabla\cdot u \, , \,
  r^{-\lambda} S_{k_\pm} u|_{\varphi=\pm\theta_\xi/2}\, , \,
  r^{1-\lambda}N_{k_\pm}(u,p)|_{\varphi=\pm\theta_\xi/2}\big),
\]
where $u(x)=r^\lambda U(\varphi)$, $p(x)=r^{\lambda-1}P(\varphi)$,
$\lambda\in {\Bbb C}$. The operator $A_\xi(\lambda)$ depends
quadratically on the parameter $\lambda$ and realizes a continuous
mapping
\[
W^{2,2}(I_\xi)^3 \times W^{1,2}(I_\xi) \to W^{1,2}(I_\xi)^3 \times
L_2(I_\xi) \times {\Bbb C}^3\times {\Bbb C}^3
\]
for every $\lambda \in {\Bbb C}$, where $I_\xi$ denotes the
interval $(-\theta_\xi/2,+\theta_\xi/2)$. The spectrum of the
pencil $A_\xi(\lambda)$ consists of eigenvalues with finite
geometric and algebraic multiplicities. These eigenvalues are
zeros of certain transcendental functions (see \cite{mr-04a}). For
example, in the cases $d_{k_+}=d_{k_-}=0$ (Dirichlet conditions on
$\Gamma_{k_\pm}$) and $d_{k_+}=d_{k_-}=3$ (Neumann conditions on
$\Gamma_{k_\pm}$), the spectrum of $A_\xi(\lambda)$ consists of
the solutions of the equation
\[
\sin(\lambda\theta_\xi)\, \big(
\lambda^2\sin^2\theta_\xi-\sin^2(\lambda\theta_\xi)\big)=0,
\]
$\lambda\not=0$ for $d_{k_+}=d_{k_-}=0$.

Let $\lambda_1(\xi)$ be the eigenvalue with smallest positive real
part of this pencil, and let $\lambda_2(\xi)$ be the eigenvalue
with smallest real part greater than 1. We define
\[
\mu(\xi) = \left\{ \begin{array}{ll}  \mbox{Re}\, \lambda_2(\xi) &
\mbox{if }
   d_{k_+} + d_{k_-} \mbox{ is even and } \theta_\xi < \pi/m_k, \\
   \mbox{Re}\, \lambda_1(\xi) & \mbox{else},  \end{array}\right.
\]
where $m_k=1$ if $d_{k_+} + d_{k_-}\in \{ 0,6\}$, $m_k=2$ if
$d_{k_+} + d_{k_-}\in \{ 2,4\}$. Finally, let
\begin{equation} \label{i6}
\mu_k = \inf_{\xi\in M_k} \mu(\xi).
\end{equation}
Note that in the case of even $d_{k_+}+d_{k_-}$, the number
$\lambda=1$ belongs always to
the spectrum of the pencil $A_\xi(\lambda)$.\\

2) Let $x^{(j)}$ be a vertex of ${\cal G}$ and let $I_j$ be the
set of all indices $k$ such that $x^{(j)}\in \ol{\Gamma}_k$. By
our assumptions, there exist a neighborhood ${\cal U}$ of
$x^{(j)}$ and a diffeomorphism $\kappa$ mapping ${\cal G}\cap
{\cal U}$ onto ${\cal K}_j\cap B_1$ and $\Gamma_k\cap {\cal U}$
onto $\Gamma_k^\circ\cap B_1$ for $k\in I_j$, where ${\cal
K}_j=\{x:\, x/|x|\in \Omega_j\}$ is a polyhedral cone with vertex
$0$ and $\Gamma_k^\circ=\{ x:\, x/|x|\in \gamma_k\}$ are the faces
of this cone. Without loss of generality, we may assume that the
Jacobian matrix $\kappa'(x)$ is equal to the identity matrix $I$
at the point $x^{(j)}$. We introduce spherical coordinates
$\rho=|x|$, $\omega=x/|x|$ in ${\cal K}_j$ and define
\[
V_{\Omega_j} = \{ u\in W^{1,2}(\Omega_j)^3:\ S_k u=0\mbox{ on
}\gamma_k, \ k\in I_j\}.
\]
On the space $V_{\Omega_j}\times L_2(\Omega_j)$, we define the
bilinear form $a_j(\cdot,\cdot;\lambda)$ as
\[
a_j\Big( \Big( \begin{array}{c} u \\ p\end{array}\Big),\Big(
\begin{array}{c} v \\ q\end{array}\Big);\lambda\Big)
  = \frac{1}{\log 2}\, \int\limits_{\substack{{\cal K}\\ 1<|x|<2}} \Big( 2\nu\sum_{i,j=1}^3 \varepsilon_{i,j}(U)
  \cdot \varepsilon_{i,j}(V) - P\nabla\cdot V - (\nabla\cdot U)\, Q\Big)\, dx,
\]
where $U=\rho^\lambda u(\omega)$, $V=\rho^{-1-\lambda} v(\omega)$,
$P=\rho^{\lambda-1} p(\omega)$, $Q=\rho^{-2-\lambda}q(\omega)$,
$u,v \in V_{\Omega_j}$, $p,q\in L_2(\Omega_j)$, and $\lambda \in
{\Bbb C}$. This bilinear form  generates the linear and continuous
operator
\[
{\mathfrak A}_j(\lambda):\, V_{\Omega_j} \times L_2(\Omega_j) \to
V_{\Omega_j}^*\times L_2(\Omega_j)
\]
by
\[
\int_\Omega {\mathfrak A}_j(\lambda)\Big( \begin{array}{c} u \\
p\end{array}\Big) \cdot
  \Big( \begin{array}{c} v \\ q\end{array}\Big)\, d\omega
  = a_j\Big( \Big( \begin{array}{c} u \\ p\end{array}\Big),\Big( \begin{array}{c} v \\ q\end{array}\Big)
  ;\lambda\Big), \quad   u,v\in V_{\Omega_j},\ p,q \in L_2(\Omega_j).
\]
The operator ${\mathfrak A}_j(\lambda)$ depends quadratically on
the complex parameter $\lambda$. The spectrum of the pencil
${\mathfrak A}_j(\lambda)$ consists of isolated points,
eigenvalues with finite geometric and algebraic multiplicities.

\subsection{Regularity assertions for weak solutions of the linearized problem}

The following two theorems are proved in \cite{mr-04b}.

\begin{Th} \label{Stokest2a}
Let $(u,p)\in W^{1,2}({\cal G})^3\times L_2({\cal G})$ be a
solution of the problem {\em (\ref{Stokes1}), (\ref{Stokes2})}.
Suppose that the following conditions are satisfied.
\begin{itemize}
\item[{\em (i)}] $F\in V^* \cap V_{\beta,\delta}^{-1,s}({\cal G})^3$, \ \  $g\in L_2({\cal G})
  \cap W_{\beta,\delta}^{0,s}({\cal G})$, \ \ $h_j \in W^{1/2,2}(\Gamma_j) \cap W^{1-1/s,s}_{\beta,\delta}(\Gamma_j)$,
\item[{\em (ii)}] there are no eigenvalues of the pencils ${\mathfrak A}_j(\lambda)$, $j=1,\ldots,d$ in the closed
  strip between the lines $\mbox{\em Re}\lambda = -1/2$ and $\mbox{\em Re}\lambda = 1-\beta_j-3/s$,
\item[{\em (iii)}] the components of $\delta$ satisfy the inequalities $\max(1-\mu_k,0)<\delta_k+2/s<1$.
\end{itemize}
Then $u\in W^{1,s}_{\beta,\delta}({\cal G})^3$ and $p\in
W_{\beta,\delta}^{0,s}({\cal G})$.
\end{Th}

\begin{Th} \label{Stokest2b}
Let $(u,p)\in W^{1,2}({\cal G})^3\times L_2({\cal G})$ be a
solution of the problem {\em (\ref{Stokes1}), (\ref{Stokes2})}.
Suppose that
\begin{itemize}
\item[{\em (i)}] $g\in W_{\beta,\delta}^{1,s}({\cal G})$, $h_j\in W_{\beta,\delta}^{2-1/s,s}(\Gamma_j)^{3-d_j}$, and
$F\in V^*$ has the representation {\em (\ref{a4})} with $f\in
W_{\beta,\delta}^{0,s}({\cal G})^3$, $\phi_j \in
W_{\beta,\delta}^{1-1/s}(\Gamma_j)^{d_j}$,
\item[{\em (ii)}] there are no eigenvalues of the pencils ${\mathfrak A}_j(\lambda)$, $j=1,\ldots,d$, in
the closed strip between the lines $\mbox{\em Re}\, \lambda= -1/2$
and $\mbox{\em Re}\, \lambda =2-\beta_j-3/s$,
\item[{\em (iii)}] the components of $\delta$ satisfy the inequalities $\max(2-\mu_k,0)<\delta_k+2/s<2$,
\item[{\em (iv)}] $g$, $h_j$ and $\phi_j$ are such that there exist $w\in W_{\beta,\delta}^{2,s}
({\cal G})^3$ and $q\in W_{\beta,\delta}^{1,s}({\cal G})$
satisfying
\[
S_jw=h_j, \ \ N_j(w,q)=\phi_j \ \mbox{ on }\Gamma_j,\
j=1,\ldots,n,\quad
  \nabla\cdot w +g \in V_{\beta,\delta}^{1,s}({\cal G}).
\]
\end{itemize}
Then $u\in W_{\beta,\delta}^{2,s}({\cal G})^3$ and $p\in
W_{\beta,\delta}^{1,s}({\cal G})$.
\end{Th}

For the proof of the following two regularity assertions in
weighted H\"older spaces, we refer to \cite{mr-05}.

\begin{Th} \label{Stokest3a}
Let $(u,p)\in W^{1,2}({\cal G})^3\times L_2({\cal G})$ be a weak
solution of the problem {\em (\ref{Stokes}), (\ref{NS2})}. Suppose
that
\begin{itemize}
\item[{\em (i)}] $f\in C_{\beta,\delta}^{-1,\sigma}({\cal G})^3$, $g\in C_{\beta,\delta}^{0,\sigma}({\cal G})$,
  $h_j\in C_{\beta,\delta}^{1,\sigma}(\Gamma_j)^{3-d_j}$, $\phi_j \in C_{\beta,\delta}^{0,\sigma}(\Gamma_j)^{d_j}$,
\item[{\em (ii)}] $\beta_j-\sigma<3/2$ for $j=1,\ldots,d$, and the strip $-1/2< \mbox{\em Re}\, \lambda
  \le 1+\sigma-\beta_j$ is free of eigenvalues of the pencils ${\mathfrak A}_j(\lambda)$, $j=1,\ldots,d$,
\item[{\em (iii)}] the components of $\delta$ are nonnegative and satisfy the inequalities $1-\mu_k<\delta_k-\sigma<1$,
  $\delta_k\not=\sigma$,
\item[{\em (iv)}] $g$, $h_j$ and $\phi_j$ are such that there exist $w\in C_{\beta,\delta}^{1,\sigma}({\cal G})^3$
  and $q\in C_{\beta,\delta}^{1,\sigma}({\cal G})$ satisfying
\[
S_jw=h_j, \ \ N_j(w,q)=\phi_j \ \mbox{ on }\Gamma_j,\
j=1,\ldots,n,\quad
  \nabla\cdot w + g \in {\cal N}_{\beta,\delta}^{0,\sigma}({\cal G}).
\]
\end{itemize}
Then $(u,p)\in C_{\beta,\delta}^{1,\sigma}({\cal G})^3\times
C_{\beta,\delta}^{0,\sigma}({\cal G})$.
\end{Th}

Note that under the conditions of Theorem \ref{Stokest3a} on
$\beta$ and $\delta$, there are the imbeddings
$C_{\beta,\delta}^{-1,\sigma}({\cal G})^3 \subset V^*$,
$C_{\beta,\delta}^{0,\sigma}({\cal G}) \subset L_2({\cal G})$, and
$C_{\beta,\delta}^{1,\sigma}(\Gamma_j) \subset W^{1/2,2}({\cal
G})$.

\begin{Th} \label{Stokest3}
Let $(u,p)\in W^{1,2}({\cal G})^3\times L_2({\cal G})$ be a weak
solution of the problem {\em (\ref{Stokes}), (\ref{NS2})}. Suppose
that
\begin{itemize}
\item[{\em (i)}] $f\in C_{\beta,\delta}^{0,\sigma}({\cal G})^3$, $g\in C_{\beta,\delta}^{1,\sigma}({\cal G})$,
  $h_j\in C_{\beta,\delta}^{2,\sigma}(\Gamma_j)^{3-d_j}$, $\phi_j \in C_{\beta,\delta}^{1,\sigma}(\Gamma_j)^{d_j}$,
\item[{\em (ii)}] $\beta_j-\sigma<5/2$ for $j=1,\ldots,d$, the strip $-1/2<\mbox{\em Re}\, \lambda\le 2+\sigma-\beta_j$
  is free of eigenvalues the pencil ${\mathfrak A}_j(\lambda)$, $j=1,\ldots,d$,
\item[{\em (iii)}] the components of $\delta$ are nonnegative, satisfy the inequalities $2-\mu_k<\delta_k-\sigma<2$,
$\delta_k\not=\sigma$, and $\delta_k\not=1+\sigma$,
\item[{\em (iv)}] $g$, $h_j$ and $\phi_j$ are such that there exist $w\in C_{\beta,\delta}^{2,\sigma}({\cal G})^3$
  and $q\in C_{\beta,\delta}^{1,\sigma}({\cal G})$ satisfying
\[
S_jw=h_j, \ \ N_j(w,q)=\phi_j \ \mbox{ on }\Gamma_j,\
j=1,\ldots,n,\quad
  \nabla\cdot w + g \in {\cal N}_{\beta,\delta}^{1,\sigma}({\cal G}).
\]
\end{itemize}
Then $(u,p)\in C_{\beta,\delta}^{2,\sigma}({\cal G})^3\times
C_{\beta,\delta}^{1,\sigma}({\cal G})$.
\end{Th}

\begin{Rem}
{\em For the validity of condition (iv) in Theorems
\ref{Stokest2b}--\ref{Stokest3} it is necessary and sufficient
that the functions $h_j$ and their derivatives, $\phi_j$ and $g$
satisfy certain compatibility conditions on the edges of the
domain ${\cal G}$ (see \cite{mr-04b,mr-05}).}
\end{Rem}

\subsection{Regularity results for solutions of the nonlinear problem in weighted Sobolev spaces}

Our goal is to extend the results of Theorems
\ref{Stokest2a}--\ref{Stokest3} to the nonlinear problem
(\ref{NS1}), (\ref{NS2}). We start with regularity results in
weighted Sobolev spaces.

\begin{Le} \label{dl1}
Let $u\in L_6({\cal G})\cap W_{\beta,\delta}^{1,s}({\cal G})$, $s>
6/5$, $\beta'_j \ge  \beta_j -1/2$ for $j=1,\ldots,d$, and
$\delta'_k \ge \delta_k -1/2$. Then $u \, \partial_{x_i} u \in
V_{\beta',\delta'}^{-1,s}({\cal G})$ for $i=1,2,3$.
\end{Le}

P r o o f. Let $q=6s/(s+6)$. By H\"olders inequality,
\[
\| u\, \partial_{x_i}\, u\|_{V_{\beta,\delta}^{0,q}({\cal G})}
  \le \| u\|_{L_6({\cal G})} \ \| \partial_{x_i}u\|_{V_{\beta,\delta}^{0,s}({\cal G})}.
\]
Furthermore by Corollary \ref{ac1}, the space
$V_{\beta,\delta}^{0,q}({\cal G})$ is continuously imbedded into
$V_{\beta',\delta'}^{-1,s}({\cal G})$ if $\beta'_j\ge \beta_j-1/2$
and $\delta'_k\ge \delta_k-1/2$. The result follows. \hfill $\Box$

\begin{Th} \label{dt1}
Let $(u,p)\in W^{1,2}({\cal G})^3\times L_2({\cal G})$ be a
solution of the problem {\em (\ref{a1}), (\ref{a2})}. We suppose
that $s>6/5$ and that the conditions {\em (i)-(iii)} of Theorem
{\em \ref{Stokest2a}} are satisfied. Then $u \in
W_{\beta,\delta}^{1,s}({\cal G})^3$ and $p\in
W_{\beta,\delta}^{0,s}({\cal G})$.
\end{Th}

P r o o f. 1) First, let $s\le 3$. From $u_j\in W^{1,2}({\cal
G})\subset L_6({\cal G})$ and $\partial_{x_j}u\in L_2({\cal G})^3$
it follows that $u_j\, \partial_{x_j}u \in L_{3/2}({\cal G})^3$.
This together with Corollary \ref{ac1} implies $(u\cdot\nabla)\, u
\in V_{1-3/s,1-3/s}^{-1,s}({\cal G})^3$. Hence, $(u,p)$ is a
solution of the problem
\begin{eqnarray} \label{1dt1}
&& b(u,v)-\int_{\cal G} p\, \nabla\cdot v\, dx = \Phi(v)\ \mbox{
for all }v\in V, \\ \label{2dt1} && -\nabla\cdot u = g \ \mbox{ in
}{\cal G},\quad S_j u =h_j\ \mbox{ on }\Gamma_j,\ j=1,\ldots,N,
\end{eqnarray}
where
\[
\Phi = F - (u\cdot \nabla)u \in V_{\beta',\delta'}^{-1,s}({\cal
G})^3, \quad \beta'_j=\max(\beta_j,1-3/s), \
  \delta'_k=\max(\delta_k,1-3/s).
\]
From Theorem \ref{Stokest2a} we conclude that $(u,p)\in
W_{\beta',\delta'}^{1,s}({\cal G})^3 \times
W_{\beta',\delta'}^{0,s}({\cal G})$. Then by Lemma \ref{dl1}, we
have $(u\cdot \nabla)\, u \in
V_{\beta'-1/2,\delta'-1/2}^{-1,s}({\cal G})^3$ and therefore,
\[
F-(u\cdot \nabla)u \in V_{\beta'',\delta''}^{-1,s}({\cal G})^3, \
\ \mbox{ where }
 \beta''_j=\max(\beta_j,1/2-3/s), \ \delta''_k=\max(\delta_k,1/2-3/s).
\]
Consequently, Theorem \ref{Stokest2a} implies $(u,p)\in
W_{\beta'',\delta''}^{1,s}({\cal G})^3 \times
W_{\beta'',\delta''}^{0,s}({\cal G})$. Repeating the last
consideration, we obtain $(u,p)\in W_{\beta,\delta}^{1,s}({\cal
G})^3 \times W_{\beta,\delta}^{0,s}({\cal G})$.

2) Next, let $3<s\le 6$. Then by Lemma \ref{al7},
$V_{\beta,\delta}^{-1,s}({\cal G}) \subset
V_{\beta',\delta'}^{-1,3}({\cal G})$,
$W_{\beta,\delta}^{0,s}({\cal G}) \subset
W_{\beta',\delta'}^{0,3}({\cal G})$, and
$W_{\beta,\delta}^{1-1/s,s}(\Gamma_j) \subset
W_{\beta',\delta'}^{2/3,3}(\Gamma_j)$, where
$\beta'_j=\beta+3/s-1+\varepsilon$,
$\delta'_k=\delta_k+2/s-2/3+\varepsilon$, $\varepsilon$ is an
arbitrarily small positive number. For sufficiently small
$\varepsilon$, we conclude from part 1) that $u\in
W_{\beta',\delta'}^{1,3}({\cal G})^3$. By H\"olders inequality,
\[
\| u_j\, \partial_{x_j} u\|_{W_{\beta',\delta'}^{0,2}({\cal G})^3}
\le
  \| u_j\|_{L_6({\cal G})} \ \| \partial_{x_j} u\|_{W_{\beta',\delta'}^{0,3}({\cal G})^3}
\]
Due to Corollary \ref{ac1}, $W_{\beta',\delta'}^{0,2}({\cal G})
\subset V_{\beta,\delta}^{-1,s}({\cal G})$ if $\varepsilon<1/3$.
Therefore, $(u,p)$ is a solution of the problem (\ref{1dt1}),
(\ref{2dt1}), where $\Phi=F - (u\cdot\nabla)\, u \in
V_{\beta,\delta}^{-1,s}({\cal G})^3$. Using Theorem
\ref{Stokest2a}, we obtain $(u,p) \in W_{\beta,\delta}^{1,s}({\cal
G})^3\times W_{\beta,\delta}^{0,s}({\cal G})$.

3) Finally, let $s>6$. Then again by Lemma \ref{al7},
$V_{\beta,\delta}^{-1,s}({\cal G}) \subset
V_{\beta',\delta'}^{-1,6}({\cal G})$,
$W_{\beta,\delta}^{0,s}({\cal G}) \subset
W_{\beta',\delta'}^{0,6}({\cal G})$, and
$W_{\beta,\delta}^{1-1/s,s}(\Gamma_j) \subset
W_{\beta',\delta'}^{5/6,6}(\Gamma_j)$, where
$\beta'_j=\beta+3/s-1/2+\varepsilon$,
$\delta'_k=\delta_k+2/s-1/3+\varepsilon$, $\varepsilon$ is an
arbitrarily small positive number. For sufficiently small
$\varepsilon$, we conclude from part 2) that $u\in
W_{\beta',\delta'}^{1,6}({\cal G})^3$. Since $u\in L_6({\cal
G})^3$, it follows that $(u\cdot\nabla)\, u \in
W_{\beta',\delta'}^{0,3}({\cal G})^3$. The last space is embedded
to $V_{\beta,\delta}^{-1,s}({\cal G})^3$ if $\varepsilon \le 1/3$
(see Corollary \ref{ac1}). Therefore, $(u,p)$ is a solution of the
problem (\ref{1dt1}), (\ref{2dt1}), where $\Phi\in
V_{\beta,\delta}^{-1,s}({\cal G})^3$. Applying Theorem
\ref{Stokest2a}, we obtain
$(u,p) \in W_{\beta,\delta}^{1,s}({\cal G})^3\times W_{\beta,\delta}^{0,s}({\cal G})$. \hfill $\Box$ \\

For the proof of the analogous $W_{\beta,\delta}^{2,s}$ regularity
result, we need the following lemma.

\begin{Le} \label{dl2}
Let $u\in W_{\beta,\delta}^{2,s}({\cal G})\cap W^{1,2}({\cal G})$,
$1<s<6$, $\beta_j+3/s \le 5/2$, $\delta_k+2/s>0$. Then $u\, \nabla
u\in V_{\beta-1/2,\delta'}^{0,s}({\cal G})^3$ for every $\delta'$,
$\delta'_k\ge \delta_k-1/2$, $\delta'_k+2/s >0$.
\end{Le}

P r o o f. 1) Suppose that $\delta_k +2/s >1$ for $k=1,\ldots,m$.
Then by Lemma \ref{al8}, $\nabla u\in W_{\beta,\delta}^{1,s}({\cal
G})^3 = V_{\beta,\delta}^{1,s}({\cal G})^3 \subset
V_{\beta-1/2,\delta-1/2}^{0,q}({\cal G})^3$, $q=6s/(6-s)$. From
this, from the assumption $u\in W^{1,2}({\cal G})\subset L_6({\cal
G})$ and from H\"older's inequality it follows that $u\, \nabla u
\in V_{\beta-1/2,\delta-1/2}^{0,s}({\cal G})^3$,
\[
\| u\, \nabla u \|_{V_{\beta-1/2,\delta-1/2}^{0,s}({\cal G})^3}
\le c\, \| u\|_{L_6({\cal G})}
  \ \| \nabla u \|_{V_{\beta-1/2,\delta-1/2}^{0,q}({\cal G})^3}
\]
2) We consider the case when $0 < \delta_k+2/s \le 1$ for all $k$.
Then $u$ admits the decomposition
\[
u= v + w,\quad v\in V_{\beta,\delta}^{2,s}({\cal G}), \ \ w\in
W_{\beta+2,\delta+2}^{4,s}({\cal G}).
\]
From Lemma \ref{al3} it follows that
\begin{equation} \label{1dl2}
\prod_j\rho_j^{\beta_j-1+2/s} \, \nabla w \in L_{3s}({\cal
G})^3,\quad \prod_j\rho_j^{\beta_j-2+3/s}w \in L_\infty({\cal
G}),\quad \prod_j\rho_j^{\beta_j-5/2+3/s}w \in L_6({\cal G}).
\end{equation}
Since $\beta_j-5/2+3/2\le 0$, we have in particular $w\in
L_6({\cal G})$ and therefore also $v\in L_6({\cal G})$. We
estimate the norms of $v\nabla u$ and $w\nabla u$ in
$V_{\beta-1/2,\delta-1/2}^{0,s}({\cal G})^3$. Let $q=6s/(6-s)$.
Using H\"older's inequality and Lemma \ref{al8}, we obtain
\[
\| v\nabla v\|_{V_{\beta-1/2,\delta-1/2}^{0,s}({\cal G})^3} \le
\|v\|_{L_6({\cal G})} \
  \| \nabla v\|_{V_{\beta-1/2,\delta-1/2}^{0,q}({\cal G})^3} \le c\, \|v\|_{L_6({\cal G})} \
  \| \nabla v\|_{V_{\beta,\delta}^{1,s}({\cal G})^3}\, .
\]
By Lemma \ref{al8}, the space $V_{\beta,\delta}^{2,s}({\cal G})$
is continuously imbedded into
$V_{\beta-2+1/s,\delta-2+1/s}^{0,3s/2}({\cal G})$. Consequently,
\begin{eqnarray*}
\| v\, \nabla w\|_{V_{\beta-1/2,\delta-1/2}^{0,s}({\cal G})^3} &
\le &
  \| v\|_{V_{\beta-2+1/s,\delta-2+1/s}^{0,3s/2}({\cal G})} \
  \Big\| \prod_j\rho_j^{3/2-1/s} \, \prod_k \big( \frac{r_k}\rho \big)^{3/2-1/s}\, \nabla w \Big\|_{L_{3s}({\cal G})^3}
\\ & \le & c\, \| v\|_{V_{\beta,\delta}^{2,s}({\cal G})} \
  \Big\| \prod_j\rho_j^{\beta_j-1+2/s} \, \nabla w \Big\|_{L_{3s}({\cal G})^3}\, .
\end{eqnarray*}
Thus, we have $v\nabla u \in V_{\beta-1/2,\delta-1/2}^{0,s}({\cal
G})^3 \subset V_{\beta-1/2,\delta'}^{0,s}({\cal G})^3$.
Furthermore, using the continuity of the imbedding
$W_{\beta,\delta}^{2,s}({\cal G}) \subset
W_{3/2-3/s,\delta'}^{1,s}({\cal G})$, we obtain
\begin{eqnarray*}
\| w\, \nabla u\|_{V_{\beta-1/2,\delta'}^{0,s}({\cal G})^3} & \le
& \Big\| \prod_j
  \rho_j^{\beta_j-2+3/s}w \Big\|_{L_\infty({\cal G})} \
  \Big\| \prod_j \rho_j^{3/2-3/s} \prod_k \big( \frac{r_k}\rho\big)^{\delta'_k}\, \nabla u\Big\|_{L_s({\cal G})^3}\\
& \le & c\, \Big\| \prod_j \rho_j^{\beta_j-2+3/s}w
\Big\|_{L_\infty({\cal G})}
   \| u\|_{W_{\beta,\delta}^{2,s}({\cal G})^3}\, .
\end{eqnarray*}
This proves the lemma for the case $\delta_k+2/s <1$,
$k=1,\ldots,m$.

3) The case when $\delta_k+2/s <1$ for some but not all $k$ can be
reduced to cases 1) and 2) using suitable cut-off functions.
\hfill $\Box$

\begin{Th} \label{dt2}
Let $(u,p)\in W^{1,2}({\cal G})^3\times L_2({\cal G})$ be a
solution of the problem {\em (\ref{a1}), (\ref{a2})}. We assume
that the conditions {\em (i)--(iv)} of Theorem {\em
\ref{Stokest2b}} are satisfied and that $\beta_j+3/s<5/2$ for
$j=1,\ldots,d$. Then $u\in W_{\beta,\delta}^{2,s}({\cal G})^3$ and
$p\in W_{\beta,\delta}^{1,s}({\cal G})$.
\end{Th}

P r o o f. 1) Let first $1<s\le 3/2$. We put $q=3s/(3-2s)$ if
$s<3/2$, $q=\infty$ if $s=3/2$. Since
\[
\| u_i \partial_{x_i} u\|_{W_{\beta',\delta'}^{0,s}({\cal G})^3}
\le \| u_i\|_{L_6({\cal G})}\
   \| \partial_{x_i}u \|_{L_2({\cal G})^3} \ \Big\| \prod_j \rho_j^{\beta'_j} \prod_k \big(
   \frac{r_k}\rho \big)^{\delta'_k}\Big\|_{L_q({\cal G})},
\]
we obtain $(u\cdot\nabla)\, u \in W_{\beta',\delta'}^{0,s}({\cal
G})^3$ if $\beta'_j>-3/q$ and $\delta'_k>-2/q$ or, what is the
same, if $\beta'_j+3/s >2$ for $j=1,\ldots,N$, $\delta'_k+2/s>4/3$
for $k=1,\ldots,m$. Let
$\beta'_j=\max(\beta_j,2-3/s+\varepsilon)$,
$\delta'_k=\max(\delta_k,4/3-2/s+\varepsilon)$, where
$\varepsilon$ is a sufficiently small positive number. Then
$(u,p)$ is a solution of the problem
\begin{eqnarray} \label{1dt2}
&& -\nu \Delta u + \nabla p = f',\ \ -\nabla\cdot u = g\ \mbox{ in
}{\cal G}\\ \label{2dt2} && S_j u = h_j, \ \ N_j(u,p)=\phi_j\
\mbox{ on }\Gamma_j,\ j=1,\ldots,N,
\end{eqnarray}
where $f'=f-(u\cdot\nabla)\, u \in W_{\beta',\delta'}^{0,s}({\cal
G})^3$, $g\in W_{\beta',\delta'}^{1,s}({\cal G})$, $h_j \in
W_{\beta',\delta'}^{2-1/s,s}(\Gamma_j)$ and $\phi_j \in
W_{\beta',\delta'}^{1-1/s,s}(\Gamma_j)$. Consequently by Theorem
\ref{Stokest2b}, we have $(u,p)\in W_{\beta',\delta'}^{2,s}({\cal
G})^3\times W_{\beta',\delta'}^{1,s}({\cal G})$. Applying Lemma
\ref{dl2}, we obtain $f'\in W_{\beta'',\delta''}^{0,s}({\cal
G})^3$, where $\beta''_j=\max(\beta_j,3/2-3/s+\varepsilon)$ and
$\delta''_k=\max(\delta_k,5/6-2/s+\varepsilon)$. Hence, Theorem
\ref{Stokest2b} implies $(u,p)\in W_{\beta'',\delta''}^{2,s}({\cal
G})^3\times W_{\beta'',\delta''}^{1,s}({\cal G})$. Repeating this
procedure, we obtain $(u,p)\in W_{\beta,\delta}^{2,s}({\cal
G})^3\times W_{\beta,\delta}^{1,s}({\cal G})$.

2) Next, we consider the case $3/2 < s \le 2$. Let $\varepsilon$
be a positive number less than 1/2 such that $\delta_k+2/s <
2-\varepsilon$ for all $k$. Then by Lemma \ref{al7},
$W_{\beta,\delta}^{l,s}({\cal G}) \subset
W_{\beta+3/s-2+\varepsilon,2/3-\varepsilon}^{l,3/2}({\cal G})$,
and from part 1) it follows that $(u,p) \in
W_{\beta+3/s-2+\varepsilon,2/3-\varepsilon}^{2,3/2}({\cal
G})^3\times
W_{\beta+3/s-2+\varepsilon,2/3-\varepsilon}^{1,3/2}({\cal G})$ if
$\varepsilon$ is sufficiently small. In particular,
\[
\partial_{x_i} u\in W_{\beta+3/s-2+\varepsilon,2/3-\varepsilon}^{1,3/2}({\cal G})^3
= V_{\beta+3/s-2+\varepsilon,2/3-\varepsilon}^{1,3/2}({\cal G})^3
\subset V_{\beta+3/s-2+\varepsilon,2/3-\varepsilon}^{0,3}({\cal
G})^3
\]
(see Lemma \ref{al8}). Let $\delta'_k=\max(\delta_k,5/3-2/s)$ for
$k=1,\ldots,m$. Then
\[
\| u_i\, \partial_{x_i} u\|_{W_{\beta,\delta'}^{0,s}({\cal G})^3}
\le  c\, \| u_i  \|_{L_6({\cal G})}\
  \Big\| \prod_j \rho_j^{\beta_j+3/s-2+\varepsilon}\, \prod_k \big(
  \frac{r_k}\rho \big)^{2/3-\varepsilon} \partial_{x_i}u\Big\|_{L_3({\cal G})^3}\, ,
\]
where
\[
c= \Big\| \prod_j \rho_j^{2-3/s-\varepsilon}\, \prod_k \big(
  \frac{r_k}\rho \big)^{\delta'_k-2/3+\varepsilon} \Big\|_{L_{2s/(2-s)}({\cal G})} <\infty.
\]
Consequently, $f'=f-(u\cdot\nabla)\, u \in
W_{\beta,\delta'}^{0,s}({\cal G})^3$, and Theorem \ref{Stokest2b}
implies $(u,p)\in W_{\beta,\delta'}^{2,s}({\cal G})^3\times
W_{\beta,\delta'}^{1,s}({\cal G})$. Using Lemma \ref{dl2}, we
obtain $(u,p)\in W_{\beta,\delta}^{2,s}({\cal G})^3\times
W_{\beta,\delta}^{1,s}({\cal G})$ analogously to the first part of
the proof.

3) Let $2<s\le 3$, and let $\varepsilon$ be a positive number less
than 1/2 such that $\delta_k+2/s < 2-\varepsilon$ for all $k$.
Then by Lemma \ref{al7}, $W_{\beta,\delta}^{l,s}({\cal G}) \subset
W_{\beta+3/s-3/2+\varepsilon,1-\varepsilon}^{l,2}({\cal G})$.
Therefore by part 2), we have $(u,p) \in
W_{\beta+3/s-3/2+\varepsilon,1-\varepsilon}^{2,2}({\cal
G})^3\times
W_{\beta+3/s-3/2+\varepsilon,1-\varepsilon}^{1,2}({\cal G})$
provided $\varepsilon$ is sufficiently small. Consequently,
\[
\partial_{x_i} u\in W_{\beta+3/s-3/2+\varepsilon,1-\varepsilon}^{1,2}({\cal G})^3
= V_{\beta+3/s-3/2+\varepsilon,1-\varepsilon}^{1,2}({\cal G})^3
\subset V_{\beta+3/s-3/2+\varepsilon,1-\varepsilon}^{0,6}({\cal
G})^3.
\]
Let $\delta'_k=\max(\delta_k,5/3-2/s)$ for $k=1,\ldots,m$. Then
\[
\| u_i\, \partial_{x_i} u\|_{W_{\beta,\delta'}^{0,s}({\cal G})^3}
\le  c\,  \| u_i \|_{L_6({\cal G})}\
  \| \partial_{x_i} u \|_{V_{\beta+3/s-3/2+\varepsilon,1-\varepsilon}^{0,6}({\cal G})^3}
  \, ,
\]
where
\[
c = \Big\| \prod_j \rho_j^{3/2-3/s-\varepsilon}\, \prod_k \big(
  \frac{r_k}\rho \big)^{\delta'_k-1+\varepsilon} \Big\|_{L_{3s/(3-s)}({\cal G})} < \infty .
\]
Thus, we have $f'=f-(u\cdot\nabla)\, u \in
W_{\beta,\delta'}^{0,s}({\cal G})^3$ and $(u,p)\in
W_{\beta,\delta'}^{2,s}({\cal G})^3\times
W_{\beta,\delta'}^{1,s}({\cal G})$ (by Theorem \ref{Stokest2b}).
Using again Lemma \ref{dl2}, we obtain $(u,p)\in
W_{\beta,\delta}^{2,s}({\cal G})^3\times
W_{\beta,\delta}^{1,s}({\cal G})$.

4) Finally, let $s>3$. We define $\delta'_k
=\max(\delta_k,1-2/s+\varepsilon)$, where $\varepsilon$ is a
sufficiently small positive number. Then we have
\[
g\in W_{\beta,\delta'}^{1,s}({\cal G})\subset
W_{\beta-1,\delta'-1}^{0,s}({\cal G}), \quad h_j \in
W_{\beta,\delta'}^{2-1/s,s}(\Gamma_j)^{3-d_j}\subset
W_{\beta-1,\delta'-1}^{1-1/s,s}(\Gamma_j)^{3-d_j}.
\]
Furthermore, the functional (\ref{a4}) belongs to
$V_{\beta-1,\delta'-1}^{-1,s}({\cal G})^3$. Since
$\max(1-\mu_k,0)<\delta'_k-1+2/s<1$ it follows from Theorem
\ref{dt1} that $u \in W_{\beta-1,\delta'-1}^{1,s}({\cal G})^3$.
Then by Corollary \ref{ac2},
\[
\prod_j \rho_j^{\beta_j-2+3/s}\, \prod_k \big( \frac{r_k}\rho
\big)^{\sigma_k} \, u \in L_\infty({\cal G})^3,
\]
where $\sigma_k=0$ for $\delta_k<2-3/s$,
$\sigma_k=1/s+\varepsilon$ for $2-3/s \le \delta_k<2-2/s$. By
H\"older's inequality, we have
\begin{eqnarray*}
\| u_i\, \partial_{x_i}u\|_{W_{\beta,\delta}^{0,s}({\cal G})} &
\le &
  \Big\| \prod_j \rho_j^{\beta_j-2+3/s}\, \prod_k \big( \frac{r_k}\rho \big)^{\sigma_k} \, u_i
  \Big\|_{L_\infty({\cal G})} \ \Big\| \prod_j \rho_j^{2-3/s}\, \prod_k \big( \frac{r_k}\rho
  \big)^{\delta_k-\sigma_k}\, \partial_{x_i}u\Big\|_{L_s({\cal G})^3} \\
& \le & c\, \Big\| \prod_j \rho_j^{\beta_j-2+3/s}\, \prod_k \big(
\frac{r_k}\rho \big)^{\sigma_k} \, u_i
  \Big\|_{L_\infty({\cal G})} \ \| \partial_{x_i} u\|_{W_{\beta-1,\delta'-1}^{0,s}({\cal G})^3}\, .
\end{eqnarray*}
For the last inequality, we used the fact that $\beta-1\le 2-3/s$
and $\delta'_k-1\le \delta_k-\sigma_k$. Hence, $(u,p)$ is a
solution of the problem (\ref{1dt1}), (\ref{2dt1}), where $\Phi =
F - (u\cdot\nabla)\, u$ is a functional of the form (\ref{a4})
with $f\in W_{\beta,\delta}^{0,2}({\cal G})$, $\phi_j \in
W_{\beta,\delta}^{1-1/s,s}({\cal G})^{d_j}$. Applying Theorem
\ref{Stokest2b}, we obtain $u\in W_{\beta,\delta}^{2,s}({\cal
G})^3$ and $p\in W_{\beta,\delta}^{1,s}({\cal G})$. \hfill $\Box$

\subsection{Regularity results in weighted H\"older spaces}

In order to extend the results of Theorems \ref{Stokest3a} and
\ref{Stokest3} to problem (\ref{NS1}), (\ref{NS2}), we consider
first the nonlinear term in the Navier-Stokes system.

\begin{Le} \label{el1}
Let $u\in C_{\beta,\delta}^{2,\sigma}({\cal G})$, where
$\beta_j\le 3+\sigma$ for $j=1,\ldots,d$, $0\le
\delta_k<2+\sigma$, $\delta_k-\sigma$ is not an integer for
$k=1,\ldots,m$. Then $u\, \partial_{x_i} u\in
C_{\beta,\delta'}^{0,\sigma}({\cal G})$ for every $\delta'$ such
that $\delta'_k\ge \max(0,\delta_k-1)$, $k=1,\ldots,m$.
\end{Le}

P r o o f. We have to show that there exists a constant $C$ such
that
\begin{eqnarray} \label{1el1}
&& \prod_j \rho_j(x)^{\beta_j-\sigma}\, \prod_k
\big(\frac{r_k}\rho\big)^{\max(\delta'_k-\sigma,0)}\,
  \big| u(x)\partial_{x_i}(x)\big| \le C, \\ \label{2el1}
&& \prod_j \rho_j(x)^{\beta_j}\, \prod_k
\big(\frac{r_k(x)}{\rho(x)}\big)^{\delta'_k}\,
  \frac{| u(x)\partial_{x_i}u(x)-u(y)\partial_{y_i}u(y)|}{|x-y|^\sigma} \le C\ \ \mbox{ for } |x-y|<r(x)/2
\end{eqnarray}
and
\begin{equation} \label{3el1}
\rho_j(x)^{\beta_j-\delta'_k}\,
  \frac{| u(x)\partial_{x_i}u(x)-u(y)\partial_{y_i}u(y)|}{|x-y|^{\sigma-\delta'_k}} \le C \ \ \mbox{ for }
  \delta'_k<\sigma, \ x,y\in {\cal G}_{j,k},\ |x-y|< \rho_j(x)/2.
\end{equation}
Here ${\cal G}_{j,k}=\{ x\in {\cal G}: \, \rho_j(x)<3\rho(x)/2,\
r_k(x)<3r(x)/2\}$. Inequality (\ref{1el1}) follows immediately
from the estimates
\begin{equation} \label{4el1}
\prod_j \rho_j(x)^{\beta_j-2-\sigma+|\alpha|}\, \prod_k
\big(\frac{r_k}\rho\big)^{\max(\delta_k-2-\sigma+|\alpha|,0)}\,
  |\partial_x^\alpha u(x)| \le \| u\|_{ C_{\beta,\delta}^{2,\sigma}({\cal G})} \ \ \mbox{ for }|\alpha|\le 2
\end{equation}
and the inequalities $\beta_j-\sigma<3$, $\delta'_k\ge
\delta_k-1$. Furthermore for $|x-y|<r(x)/2$, we have
\[
\frac{|u(x)-u(y)|}{|x-y|^\sigma} \le  |x-y|^{1-\sigma}\, \big|
\nabla u\big(x+t(y-x)\big)\big|
  \le c\, r(x)^{1-\sigma}\, \prod_j \rho_j(x)^{1+\sigma-\beta_j}\, \prod_k
  \big(\frac{r_k(x)}{\rho(x)}\big)^{-\max(\delta_k-1-\sigma,0)}\,
\]
and analogously
\[
\frac{\big|\partial_{x_i}u(x)-\partial_{y_i}u(y)\big|}{|x-y|^\sigma}
\le  c\, r(x)^{1-\sigma}\, \prod_j
  \rho_j(x)^{\sigma-\beta_j}\, \prod_k  \big(\frac{r_k(x)}{\rho(x)}\big)^{-\max(\delta_k-\sigma,0)}\, .
\]
Hence,
\begin{eqnarray*}
&& \frac{|
u(x)\partial_{x_i}u(x)-u(y)\partial_{y_i}u(y)|}{|x-y|^\sigma} \le
\frac{|u(x)-u(y)|}{|x-y|^\sigma}\,
  \big|\partial_{x_i}u(x)\big| + \frac{\big|\partial_{x_i}u(x)-\partial_{y_i}u(y)\big|}{|x-y|^\sigma}\, |u(y)|\\
&& \le  c\, r(x)^{1-\sigma}\, \prod_j
\rho_j(x)^{2+2\sigma-2\beta_j}\, \Big(
  \prod_k  \big(\frac{r_k(x)}{\rho(x)}\big)^{-2\max(\delta_k-1-\sigma,0)} +
  \prod_k  \big(\frac{r_k(x)}{\rho(x)}\big)^{-\max(\delta_k-\sigma,0)}\Big)
\end{eqnarray*}
for $|x-y|<r(x)/2$. From this estimate and from the inequalities
\[
c_1\, r(x) \le \prod_j\rho_j(x) \ \prod_k  \frac{r_k(x)}{\rho(x)}
\le c_2\, r(x),
\]
$\delta'_k+1-\sigma \ge 2\max(\delta_k-\sigma-1,0)$,
$\delta'_k+1-\sigma\ge \max(\delta_k-\sigma,0)$, and
$\beta_j-\sigma<3$, we obtain (\ref{2el1}). Analogously, we obtain
the estimate
\[
\rho_j(x)^{\beta_j-\delta'_k}\,
  \frac{| u(x)-u(y)|}{|x-y|^{\sigma-\delta'_k}}\, |\partial_{x_i}u(x)| \le C \ \ \mbox{ for }
  \delta'_k<\sigma, \ x,y\in {\cal G}_{j,k},\ |x-y|< \rho_j(x)/2.
\]
Since $\partial_{x_i}u(x) \in C_{\beta,\delta}^{1,\sigma}({\cal
G})\subset C_{\beta,\delta'+1}^{1,\sigma} ({\cal G})$, there
exists a constant $C$ such that
\[
\rho_j(x)^{\beta_j-1-\delta'_k}\,
  \frac{| \partial_{x_i}u(x)-\partial_{y_i}u(y)|}{|x-y|^{\sigma-\delta'_k}}\le C \ \ \mbox{ for }
  \delta'_k<\sigma, \ x,y\in {\cal G}_{j,k},\ |x-y|< \rho_j(x)/2.
\]
This together with (\ref{4el1}) implies
\[
\rho_j(x)^{\beta_j-\delta'_k}\,
  \frac{| \partial_{x_i}u(x)-\partial_{y_i}u(y)|}{|x-y|^{\sigma-\delta'_k}}\, |u(x)| \le C \ \ \mbox{ for }
  \delta'_k<\sigma, \ x,y\in {\cal G}_{j,k},\ |x-y|< \rho_j(x)/2.
\]
Thus, estimate (\ref{3el1}) holds. The proof is complete. \hfill
$\Box$

\begin{Th} \label{et1}
Let $(u,p)\in W^{1,2}({\cal G})^3\times L_2({\cal G})$ be a weak
solution of the problem {\em (\ref{NS1}), (\ref{NS2})}, and let
the conditions {\em (i)--(iv)} of Theorem {\em \ref{Stokest3}} are
satisfied. Then $u\in C_{\beta,\delta}^{2,\sigma}({\cal G})^3$ and
$p\in C_{\beta,\delta}^{1,\sigma}({\cal G})$.
\end{Th}

P r o o f. Suppose first that $\delta_k > \sigma$ for
$k=1,\ldots,m$. Let $\varepsilon$ be an arbitrarily small positive
number and $s$ an arbitrary real number greater than 1. We put
$\beta'_j=\beta_j-\sigma-3/s+\varepsilon$ for $j=1,\ldots,N$ and
$\delta'_k = \delta_k-\sigma-2/s+\varepsilon$. From our
assumptions on $f,g,h_j$ and $\phi_j$ it follows that
\[
f\in W_{\beta',\delta'}^{0,s}({\cal G})^3,\ \ g\in
W_{\beta',\delta'}^{1,s}({\cal G}), \ \ h_j \in
W_{\beta',\delta'}^{2-1/s,s}(\Gamma_j)^{3-d_j},\ \ \phi_j \in
W_{\beta',\delta'}^{1-1/s,s}(\Gamma_j)^{d_j}.
\]
Using Theorem \ref{dt2}, we obtain $u\in
W_{\beta',\delta'}^{2,s}({\cal G})^3$. Consequently,
\[
u \in W_{\beta'-2,-2/s+\varepsilon}^{0,s}({\cal G})^3, \ \
\partial_{x_i} u \in W_{\beta'-1,\delta''}^{0,s}({\cal G})^3, \ \ \partial_{x_i}\partial_{x_j}u
\in W_{\beta',\delta'}^{0,s}({\cal G})^3
\]
for $i,j=1,2,3$, where
$\delta''_k=\max(\delta_k-\sigma-1,0)-2/s+\varepsilon$ for
$k=1,\ldots,m$. This together with H\"older's inequality implies
\[
u_j\, \partial_{x_j}u \in
W_{2\beta'-3,\delta''-2/s+\varepsilon}^{0,s/2}({\cal G})^3,\ \
u_j\, \partial_{x_i}\partial_{x_j} u \in
W_{2\beta'-2,\delta'-2/s+\varepsilon}^{0,s/2}({\cal G})^3, \ \
(\partial_{x_i}u_j)\, \partial_{x_j} u \in
W_{2\beta'-2,2\delta''}^{0,s/2}({\cal G})^3.
\]
Therefore, $u_j\, \partial_{x_j}u \in
W_{2\beta'-2,\delta'-2/s+\varepsilon}^{1,s/2}({\cal G})^3$. The
numbers $\varepsilon$ and $s$ can be chosen such that
$\beta_j-\sigma\le 3-2\varepsilon$ for $j=1,\ldots,d$,
$\delta_k-\sigma>2/s$ for $k=1,\ldots,m$, $\varepsilon +1/s \le
1/2$, and $1-6/s>\sigma$. Then $2\beta'_j-2\le
\beta_j-\sigma+1-6/s$, $\delta'_k-2/s+\varepsilon \le
\delta_k-\sigma+1-6/s$, and $\delta_k-\sigma+1-6/s>1-4/s$.
Consequently,
\[
u_j\, \partial_{x_j}u \in
W_{\beta-\sigma+1-6/s,\delta-\sigma+1-6/s}^{1,s/2}({\cal G})^3
  = V_{\beta-\sigma+1-6/s,\delta-\sigma+1-6/s}^{1,s/2}({\cal G})^3
  \subset {\cal N}_{\beta,\delta}^{0,\sigma}({\cal G})^3
\]
(see Lemma \ref{al5}). Hence, $(u,p)$ is a solution of the problem
(\ref{1dt2}), (\ref{2dt2}), where $f'=f-(u\cdot\nabla)\, u \in
C_{\beta,\delta}^{0,\sigma}({\cal G})^3$. Applying Theorem
\ref{Stokest3}, we obtain $(u,p)\in
C_{\beta,\delta}^{2,\sigma}({\cal G})^3\times
C_{\beta,\delta}^{1,\sigma}({\cal G})$.

Suppose now that $\delta'_k<\sigma$ for at least one $k$. By the
first part of the proof, we obtain $u\in
C_{\beta,\gamma}^{2,\sigma}({\cal G})^3$, where
$\gamma_k=\max(\delta_k,\sigma+\varepsilon)$, $\varepsilon$ is an
arbitrarily small positive real number. Then Lemma \ref{el1}
implies $f'=f-(u\cdot\nabla)\, u \in
C_{\beta,\delta}^{0,\sigma}({\cal G})^3$, and from Theorem
\ref{Stokest3} it follows that $(u,p)\in
C_{\beta,\delta}^{2,\sigma}({\cal G})^3\times
C_{\beta,\delta}^{1,\sigma}({\cal G})$.
The proof is complete. \hfill $\Box$\\

Finally, we prove the analogous
$C_{\beta,\delta}^{1,\sigma}$-regularity result.

\begin{Th} \label{et2}
Let $(u,p)\in W^{1,2}({\cal G})^3\times L_2({\cal G})$ be a weak
solution of the problem {\em (\ref{NS1}), (\ref{NS2})}. Suppose
that conditions {\em (i)--(iv)} of Theorem {\em \ref{Stokest3a}}
are satisfied. Then $u\in C_{\beta,\delta}^{1,\sigma}({\cal G})^3$
and $p\in C_{\beta,\delta}^{0,\sigma}({\cal G})$.
\end{Th}

P r o o f. 1) Let first $\delta_k > \sigma$ for $k=1,\ldots,m$.
Then $g\in W_{\beta',\delta'}^{0,s}({\cal G})$ and $h_j \in
W_{\beta',\delta'}^{1-1/s,s}(\Gamma_j)^{3-d_j}$, where $\beta'_j
=\beta_j-\sigma-3/s+\varepsilon$,
$\delta'_k=\delta_k-\sigma-2/s+\varepsilon$, $\varepsilon$ is an
arbitrarily mall positive number, and $s>1$. Furthermore, the
functional (\ref{a4}) belongs to $V_{\beta',\delta'}^{-1,s}({\cal
G})^3$. Using Theorem \ref{dt1}, we obtain $(u,p)\in
W_{\beta',\delta'}^{1,s}({\cal G})^3 \times
W_{\beta',\delta'}^{1,s}({\cal G})^3$. We consider the term
\[
(u\cdot\nabla)\, u = \sum_{j=1}^3 \partial_{x_j}(u_j u) -
\sum_{j=1}^3 u\, \partial_{x_j}u_j
  = \sum_{j=1}^3 \partial_{x_j}(u_j u) + gu.
\]
From the inclusions $u_i \in
W_{\beta'-1,-2/s+\varepsilon}^{0,s}({\cal G})$, $\partial_{x_k}u_i
\in W_{\beta',\delta'}^{0,s}({\cal G})$ it follows that
\[
u_j u \in W_{2\beta'-1,\delta'-2/s+\varepsilon}^{1,s/2}({\cal
G})^3 \subset
  W_{\beta-\sigma+1-6/s,\delta-\sigma+1-6/s}^{1,s/2}({\cal G})^3
  = V_{\beta-\sigma+1-6/s,\delta-\sigma+1-6/s}^{1,s/2}({\cal G})^3
  \subset {\cal N}_{\beta,\delta}^{0,\sigma}({\cal G})^3
\]
if $\varepsilon$ is sufficiently small and $s$ is sufficiently
large (see Lemma \ref{al5}). Furthermore, from $g\in {\cal
N}_{\beta,\delta}^{0,\sigma}({\cal G})$,
\[
u \in W_{\beta',1-2/s+\varepsilon}^{1,s}({\cal G})^3 =
V_{\beta',1-2/s+\varepsilon}^{1,s}({\cal G})^3
  \subset {\cal N}_{\beta-1+\varepsilon,\sigma+\varepsilon+1/s}^{0,\sigma}({\cal G})^3
\]
and Lemma \ref{al6} it follows that
\[
gu \in {\cal
N}_{2\beta-1+\varepsilon-\sigma,\delta+\varepsilon+1/s}^{0,\sigma}({\cal
G})
  \subset {\cal N}_{\beta+1,\delta+1}^{0,\sigma}({\cal G}).
\]
Consequently, we have $(u\cdot \nabla)\, u \in
C_{\beta,\delta}^{-1,\sigma}({\cal G})^3$, and Theorem
\ref{Stokest3a} implies $(u,p) \in
C_{\beta,\delta}^{1,\sigma}({\cal G})^3 \times
C_{\beta,\delta}^{0,\sigma}({\cal G})$.

2) Suppose that $\delta_k < \sigma$ for some $k$. By the first
part of the proof, we have $(u,p) \in
C_{\beta,\gamma}^{1,\sigma}({\cal G})^3 \times
C_{\beta,\gamma}^{0,\sigma}({\cal G})$, where
$\gamma_k=\max(\delta_k,\sigma+\varepsilon)$ for $k=1,\ldots,m$,
$\varepsilon$ is an arbitrarily small positive number. In
particular, $u_j \in {\cal N}_{\beta-1,\gamma}^{0,\sigma}({\cal
G})$, $\partial_{x_j} u \in {\cal
N}_{\beta,\gamma}^{0,\sigma}({\cal G})^3$, and therefore (by Lemma
\ref{al6})
\[
u_j\partial_{x_j}u \in {\cal
N}_{2\beta-1-\sigma,2\gamma-\sigma}^{0,\sigma}({\cal G})^3.
\]
The last space is contained in ${\cal
N}_{\beta+1,\delta+1}^{0,\sigma}({\cal G})^3$ for sufficiently
small $\varepsilon$. Applying Theorem \ref{Stokest3a}, we obtain
$(u,p) \in C_{\beta,\delta}^{1,\sigma}({\cal G})^3\times
C_{\beta,\delta}^{0,\sigma}({\cal G})$. \hfill $\Box$

\subsection{Necessity of the conditions on \boldmath$\beta$ and $\delta$\unboldmath}

Let $\Lambda_j$ be the eigenvalue of the pencil ${\mathfrak
A}_j(\lambda)$ with smallest real part $> -1/2$. We show that the
inequalities
\begin{equation} \label{nec1}
\beta_j + 3/s > 2 -\mbox{Re}\, \Lambda_j,\qquad \delta_k+ 2/s >
2-\mu_k
\end{equation}
in Theorem \ref{dt2} cannot be weakened.

We assume first that $\beta_j+3/s \le 2-\mbox{Re}\, \Lambda_j$ for
some $j$ and that $\delta$ satisfies the second condition of
(\ref{nec1}). By our assumptions on the domain, there exist a
neighborhood ${\cal U}_j$ of $x^{(j)}$ and a diffeomorphism
$\kappa$ mapping ${\cal G}\cap {\cal U}_j$ onto the intersection
of a cone ${\cal K}_j$ with the unit ball such that
$\kappa'(x^{(j)})=I$. In the new coordinates $y=\kappa(x)$, the
Navier-Stokes system takes the form
\[
\sum_{i,j=1}^3 a_{i,j}(y) \, \frac{\partial^2 u}{\partial y_i\,
\partial y_j} + \sum_{i=1}^3 a_i(y)
  \, \frac{\partial u}{\partial y_j} + (u\cdot \kappa'\nabla_y)\, u + \kappa'\nabla_y p = f, \qquad
\kappa'\nabla_y\cdot u = g,
\]
where $a_{i,j}(0)=-\nu\delta_{i,j}$. Here by $\kappa'$ we mean the
matrix $\kappa'(\kappa^{-1}(y))$. We consider the functions
\[
u= \zeta(y) \, |y|^{\Lambda_j}\, \Phi(y/|y|),\quad p= \zeta(y) \,
|y|^{\Lambda_j-1}\, \Psi(y/|y|),
\]
where $(\Phi,\Psi)$ is an eigenvector of the pencil ${\mathfrak
A}_j(\lambda)$ corresponding to the eigenvalue $\Lambda_j$ and
$\zeta$ is a smooth cut-off function equal to one near the origin.
The eigenvector $(\Phi,\Psi)$ belongs to the space
$W_\gamma^{2,t}(\Omega_j)^3\times W_\gamma^{1,t} (\Omega_j)$ with
arbitrary $t$ and $\gamma$ satisfying
$\max(2-\mu_k,0)<\gamma_k+2/t<2$. Here, $W_\gamma^{l,t}(\Omega_j)$
is the closure of the set $C^\infty(\bar{\Omega}_j)$ with respect
to the norm
\[
\| u\|_{W_\gamma^{2,t}(\Omega_j)} = \Big(
\int\limits_{\substack{{\cal K}_j\\ 1/2<|x|<2}} \sum_{|\alpha|\le
l}
  \prod_k r_k^{t\delta_k}\, |\partial_x^\alpha u(x)|^t\, dx\Big)^{1/t} ,
\]
where $u$ is extended by $u(x)=u(x/|x|)$ to ${\cal K}_j$. In
particular, $\Phi \in L_\infty(\Omega_j)^3$ and $\Phi_k\,
\partial_{x_j} \Phi \in W_\delta^{0,s}(\Omega_j)^3$ for $k=1,2,3$.
Since $(\Phi,\Psi)$ is an eigenvector, the vector function
$|y|^{\Lambda_j}\, \big( \Phi(y/|y|),|y|^{-1}\Psi(y/|y|)\big)$ is
a solution of the linear Stokes system with zero right-hand sides.
From this and from the equalities $\kappa'(x^{(j)})=I$ and
$a_{i,j}(0)=-\nu\delta_{i,j}$, it follows that $f\in
W_{\beta_j,\delta}^{0,t}({\cal K}_j)^3$ and $g\in
W_{\beta_j,\delta}^{1,t}({\cal K}_j)$ if
$\beta_j+3/s>1-\mbox{Re}\, \Lambda_j$, $\beta_j+3/s>1-2\,
\mbox{Re}\,\Lambda_j$. Analogously, the corresponding boundary
data are from $W_{\beta_j,\delta}^{1-1/t,t}$ and
$W_{\beta_j,\delta}^{1-1/t,t}$, respectively. However, $u \not\in
W_{\beta_j,\delta}^{2,t}({\cal K}_j)^3$ for $\beta_j +3/s\le
2-\mbox{Re}\, \Lambda_j$. This example shows that the inequality
$\beta_j+3/s > 2-\mbox{Re}\, \Lambda_j$ cannot be weakened.

Now we show that the inequality $\delta_k+2/s>2-\mu_k$ cannot be
weakened. We assume for the sake of simplicity that the edge $M_k$
is a part of the $x_3$-axis and that the adjacent faces
$\Gamma_{k_+}$ and $\Gamma_{k_-}$ are plane. Let $\delta_k +2/s\le
2-\mu_k$, and let $\lambda_k$ be an eigenvalue of the pencil
$A_k(\lambda)$ with the real part $\mu_k$. Then we consider the
functions
\[
u(x) = \zeta(x) \, r^{\lambda_k}\, \Phi(\varphi),\qquad p(x) =
\zeta(x) \, r^{\lambda_k-1}\, \Psi(\varphi),
\]
where $r,\phi$ are the polar coordinates in the $(x_1,x_2)$-plane,
$(\Phi,\Psi)$ is an eigenvector of the pencil $A_k(\lambda)$
corresponding to the eigenvalue $\lambda_k$, and $\zeta$ is a
smooth cut-off function equal to one in a neighborhood of a
certain point $x^{(0)}\in M_k$ and equal to zero near the other
edges. Since the vector function $r^{\Lambda_k}\, \big(
\Phi(\varphi),r^{-1}\Psi(\varphi)\big)$ is a solution of the
linear Stokes system with zero right-hand sides, it follows that
\[
-\nu \Delta u +(u\cdot \nabla)\, u + \nabla p \in
W_{\beta,\delta}^{0,s}({\cal G})^3, \quad \nabla\cdot u \in
W_{\beta,\delta}^{1,s}({\cal G})
\]
if $\delta_k + 2/s > 1- \mu_k$. However, $u\not\in
W_{\beta,\delta}^{2,s}({\cal G})^3$ for $\delta_k + 2/s \le
2-\mu_k$. This means, the result of Theorem \ref{dt2} fails if
$1-\mu_k < \delta_k +2/s \le 2-\mu_k$.

Analogously, it can be shown that the inequalities for $\beta_j$
and $\delta_k$ in Theorems \ref{dt1}, \ref{et1} and \ref{et2}
cannot be weakened.

\subsection{Examples}

Here, we establish some regularity results for weak solutions in
the class of the nonweighted spaces $W^{l,s}({\cal G})$ and
$C^{l,\sigma}({\cal G})$. We assume that ${\cal G}$ is a
polyhedron with faces $\Gamma_j$, $j=1,...,N$, and edges $M_k$,
$k=1,\ldots,m$. The angle at the edge $M_k$ is denoted by
$\theta_k$. For the sake of simplicity, we restrict ourselves to
the case $g=0$ and to homogeneous boundary conditions
\begin{equation} \label{h6}
S_j u=0,\quad N_j(u,p)=0\quad\mbox{on }\Gamma_j,\ j=1,\ldots,N.
\end{equation}
Analogous results are valid for inhomogeneous boundary conditions
provided the boundary data satisfy certain compatibility
conditions on the edges. Note that there are the following
equalities
\[
W^{1,s}({\cal G})=V_{0,0}^{1,s}({\cal G}) \ \mbox{ if } s<2,\qquad
W^{1,s}({\cal G})=W_{0,0}^{1,s}({\cal G})
  \ \mbox{ if } s<3.
\]

{\em The Dirichlet problem}. For arbitrary $f\in W^{-1,2}({\cal
G})^3$, there exists a solution $(u,p) \in W^{1,2}({\cal
G})^3\times L_2({\cal G})$ of the Dirichlet problem
\[
-\nu \Delta u+(u\cdot \nabla)\, u + \nabla p = f,\ \ -\nabla u=0\
\mbox{in }{\cal G},\quad
  u=0\ \mbox{on }\Gamma_j,\ j=1,\ldots,N.
\]
(see e.g. \cite[Th.IV.2.1]{Girault}). This solution is unique for
sufficiently small $f$.

The regularity results established below are based on the
following properties of the operator pencils ${\mathfrak
A}_j(\lambda)$  (see \cite{kms} or \cite[Ch.5]{kmr2}).
\begin{itemize}
\item The strip $-1/2 \le \mbox{Re}\, \lambda \le 0$ is free of eigenvalues of the pencils
  ${\mathfrak A}_j(\lambda)$.
\item If the cone ${\cal K}_j$ is contained in a half-space, then the strip $-1/2 \le \mbox{Re}\, \lambda \le 1$
  contains only the eigenvalue $\lambda=1$ of the pencil ${\mathfrak A}_j(\lambda)$.
  This eigenvalue has only the eigenvector $(0,0,0,c)$, $c=const.$, and no generalized eigenvectors.
\item The eigenvalues of the pencil ${\mathfrak A}_j(\lambda)$ in the strip
  $-1/2 \le \mbox{Re}\, \lambda \le 1$ are real and monotonous with respect to the cone ${\cal K}_j$.
\end{itemize}
Moreover, the eigenvalues for a circular cone are solutions of a
certain transcendental equation (see \cite{kms} or \cite[Section
5.6]{kmr2}).

The numbers $\mu_k$ can be easily calculated. In the case
$\theta_k<\pi$, we have $\mu_k=\pi/\theta_k$. If $\theta_k>\pi$,
then $\mu_k$ is the smallest positive solution of the equation
\begin{equation} \label{eqmu}
\sin(\mu\theta_k)+ \mu\sin \theta_k=0.
\end{equation}
Note that $\mu_k>1/2$ for every $\theta_k<2\pi$, $\mu_k> 2/3$ if
$\theta_k < 3\, \mbox{arccos}\frac 14 \approx 1.2587\pi$,
$\mu_k>1$ if $\theta_k <\pi$, and $\mu_k> 4/3$ if $\theta_k<\frac
34 \pi$. Using these facts together with Theorems \ref{dt1} and
\ref{dt2}, we obtain the following assertions.
\begin{itemize}
\item If $f\in (W^{1,s'}({\cal G})^*)^3$, $2< s \le 3$, $s'=s/(s-1)$, then
  $(u,p) \in W^{1,s}({\cal G})^3\times L_s({\cal G})$.
  If the polyhedron ${\cal G}$ is convex, then this assertion is true for all $s>2$.
\item If $f\in W^{-1,2}({\cal G})^3 \cap L_s({\cal G})^3$,
  $1<s\le 4/3$, then $(u,p)\in W^{2,s}({\cal G})^3\times W^{1,s}({\cal G})$. If $\theta_k< 3\,
  \mbox{arccos}\frac 14 \approx 1.2587\pi$ for $k=1,\ldots,m$, then this result is true for $1<s\le 3/2$.
  If ${\cal G}$ is convex, then this result is valid for $1<s\le 2$. If, moreover, the angles at the edges
  are less than $\frac 34 \pi$, then the result holds even for $1<s<3$.
\end{itemize}
Furthermore, the following assertion is valid.
\begin{itemize}
\item If ${\cal G}$ is convex, $f \in C^{-1,\sigma}({\cal G})$, and $\sigma$ is sufficiently small (such that
  $1+\sigma<\pi/\theta_k$ and there are no eigenvalues of the pencils ${\mathfrak A}_j$ in the strip
  $1<\mbox{Re}\, \lambda \le 1+\sigma$), then $(u,p)\in C^{1,\sigma}({\cal G})^3 \times C^{0,\sigma}({\cal G})$.
\end{itemize}
We prove the last result. Let $\varepsilon$ be a positive number,
$\varepsilon <1-\sigma$. Since $C^{-1,\sigma}({\cal G})\subset
C_{\sigma+\varepsilon,0}^{-1,\sigma}({\cal G})$, it follows from
Theorem \ref{et2} that $(u,p) \in
C_{\sigma+\varepsilon,0}^{1,\sigma}({\cal G}) \times
C_{\sigma+\varepsilon,0}^{0,\sigma}({\cal G})$ if
$\sigma<\mu_k-1$. In particular, we have $u \in
C_{\sigma+1,0}^{1,\sigma}({\cal G}) \subset C^{0,\sigma}({\cal
G})$. This implies $u_j u \in C^{0,\sigma}({\cal G})^3$. Since
$\nabla\cdot u=0$, it follows that $(u\cdot\nabla)\, u = \sum_j
\partial_{x_j} (u_j u)  \in C^{-1,\sigma}({\cal G})^3$. Thus,
$(u,p)$ satisfies the Stokes system
\[
-\nu\, \Delta u+ \nabla p = f', \quad -\nabla\cdot u =0,
\]
where $f'=f-(u\cdot\nabla) u \in C^{-1,\sigma}({\cal G}) \subset
C^{-1,\sigma}_{0,0}({\cal G})$. Hence by \cite[Th.4.3]{mr-05}, the
solution $(u,p)$ admits the decomposition
\[
\Big( \begin{array} {c} u(x) \\ p(x) \end{array}\Big) = \Big(
\begin{array} {c} 0 \\ p(x^{(k)}) \end{array}\Big)
 + \Big( \begin{array} {c} w(x) \\ q(x) \end{array}\Big)
\]
in a neighborhood of the vertex $x^{(k)}$, where $(w,q) \in
C_{0,0}^{1,\sigma}({\cal G})^3 \times C_{0,0}^{0,\sigma}({\cal
G})$. This is true for every vertex $x^{(k)}$, $k=1,\ldots,d$.
Consequently, $(u,p) \in C^{1,\sigma}({\cal G})^3\times C^{0,\sigma}({\cal G})$. \\

For special domains, it is possible to obtain precise regularity
results. Let for example ${\cal G}$ have the form of steps as in
the first two pictures below with angles $\pi/2$ or $3\pi/2$ at
every edge or the form of two beams, where one lies on the other
as in the third picture. Note that the third polyhedron is not
Lipschitz.
\begin{figure}[h]
\includegraphics{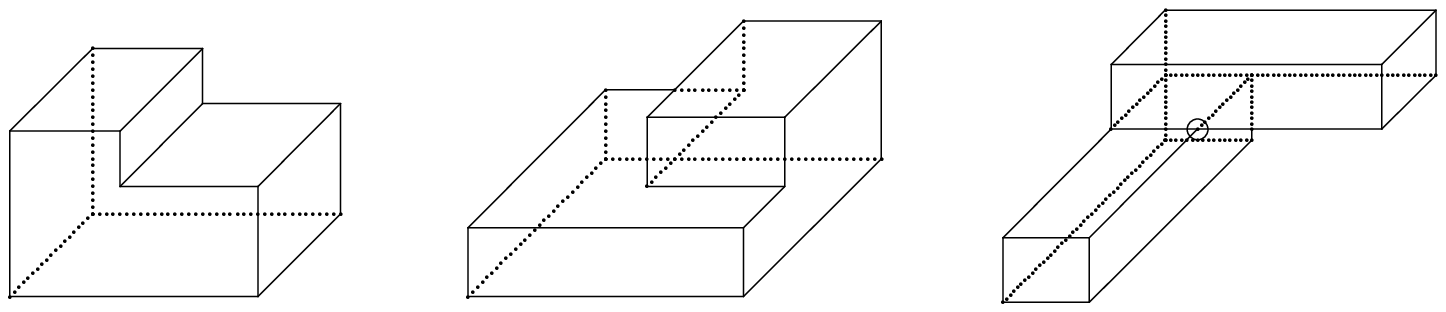}
\end{figure}
The greatest edge angle is $3\pi/2$, and we obtain $\min \mu_k =
0.54448373...$. Moreover for every vertex, there exists a circular
cone with the same vertex and aperture $3\pi/2$ which contains the
polyhedron. The left polyhedron is even contained in a half-space
bounded by a plane through an arbitrary of the vertices.
Consequently, the smallest positive eigenvalue of the pencils
${\mathfrak A}_j(\lambda)$ does not exceed the first eigenvalue
for a circular cone with vertex $3\pi/2$. A numerical calculation
shows that this eigenvalue is greater than $(3\min\mu_k -1)/2$.
This means that for $\beta=0$ and $\delta=0$, the condition (iii)
in Theorems \ref{dt1} and \ref{dt2} is stronger than the condition
(ii) in the same theorems. Thus, we obtain
\begin{itemize}
\item[] $(u,p) \in W^{1,s}({\cal G})^3 \times L_s({\cal G})$ \qquad if $f\in W^{-1,s}({\cal G})$,
  $s<2/(1-\min\mu_k)=4.3905...$,
\item[] $(u,p) \in W^{2,s}({\cal G})^3 \times W^{1,s}({\cal G})$ \qquad if $f\in L_s({\cal G})$,
  $s<2/(2-\min\mu_k)=1.3740...$.
\end{itemize}
Here the condition on $s$ is sharp.

We give some comments concerning the examples in the introduction
(the flow outside a regular polyhedron $G$). By Theorem \ref{dt2},
the regularity result $(u,p) \in W^{2,s}\times W^{1,s}$ in an
arbitrary bounded subdomain of the complement of $G$ holds if
$s<2/(2-\mu_k)$ and there are no eigenvalues of the pencils
${\mathfrak A}_j(\lambda)$ in the strip $-1/2 < \mbox{Re}\,
\lambda < 2-3/s$. Here, $\mu_k$ is the smallest positive solution
of the equation (\ref{eqmu}), where $\theta_k=\theta$ is the edge
angle in the exterior of $G$, $\sin\theta$ is equal to
$-\frac{2}{3}\sqrt{2}$ if $G$ is a regular tetrahedron or
octahedron, $-1$ if $G$ is a cube, $-\frac 25\sqrt{5}$ if $G$ is a
regular dodecahedron, and $-2/3$ if $G$ is a regular icosahedron.
The smallest positive solutions of (\ref{eqmu}) are
$\mu_k=0.52033360...$ for the regular tetrahedron,
$\mu_k=0.54448373...$ for a cube, $\mu_k=0.58489758...$ for the
regular octahedron, $\mu_k=0.60487306...$ for the regular
dodecahedron, and $\mu_k=0.68835272...$ for the regular
icosahedron. In the case of a regular tetrahedron, cube,
octahedron or dodecahedron, the inequality $s<2/(2-\mu_k)$ implies
$2-3/s< 3\mu_k/2 -1<0$. Then the absence of eigenvalues of the
pencil ${\mathfrak A}_j(\lambda)$ in the strip $-1/2 < \mbox{Re}\,
\lambda < 2-3/s$ follows from \cite[Th.5.5.6]{kmr2}. The exterior
of a regular icosahedron is contained in a right circular cone
with aperture less than $255^\circ$. Numerical results for right
circular cones  together with the monotonicity of the eigenvalues
of the pencils ${\mathfrak A}_j(\lambda)$ in the interval
$[-1/2,1)$ show that also for this polyhedron, the strip $-1/2 <
\mbox{Re}\, \lambda < 3\mu_k/2 -1$ is free of eigenvalues of the
pencil ${\mathfrak A}_j(\lambda)$. Thus, the above mentioned
regularity result holds for all $s<2/(2-\mu_k)$. This inequality
cannot be
weakened. \\

{\em The Neumann problem for the Navier-Stokes system}. We
consider a weak solution $u\in W^{1,2}({\cal G})^3\times L_2({\cal
G})$ of the Neumann problem
\[
-\Delta u+(u\cdot\nabla)\, u +\nabla p = f,\ \ -\nabla u=0\
\mbox{in }{\cal G},\quad -p n + 2\nu\varepsilon_n (u) =0\
   \mbox{on }\Gamma_j,\ j=1,\ldots,N.
\]
For this problem it is known that the strip $-1 \le \mbox{Re}\,
\lambda \le 0$ contains only the eigenvalues $\lambda=0$ and
$\lambda=1$ of the operator pencils ${\mathfrak A}_j(\lambda)$
(see \cite[Th.6.3.2]{kmr2}) if ${\cal G}$ is a Lipschitz
polyhedron. The numbers $\mu_k$ are the same as for the Dirichlet
problem. Therefore, the following assertions are valid.
\begin{itemize}
\item If $f\in (W^{1,s'}({\cal G})^*)^3$, $s'=s/(s-1)$, $2< s <3$, then $(u,p)\in W^{1,s}({\cal G})^3
  \times L_s({\cal G})$.
\item If $f\in (W^{1,2}({\cal G})^*)^3\cap L_s({\cal G})^3$, $1<s\le 4/3$, then $(u,p) \in W^{2,s}({\cal G})^3
  \times W^{1,s}({\cal G})$.If the angles $\theta_k$ are less than  $3\, \mbox{arccos}\frac 14$, then this result
  is true for $1<s< 3/2$.
\end{itemize}

{\em The mixed problem with Dirichlet and Neumann boundary
conditions}. We assume that on each face $\Gamma_j$ either the
Dirichlet condition $u=0$ or the Neumann condition $\frac{\partial
u}{\partial n}=0$ is given. If on both adjoining faces of the edge
$M_k$ the same boundary conditions are given, then $\mu_k > 1/2$.
If on one of the adjoining faces the Dirichlet condition and on
the other face the Neumann condition is given, then $\mu_k >1/4$.
This implies the following result.
\begin{itemize}
\item If $f\in (W^{1,2}({\cal G})^*)^3\cap L_s({\cal G})^3$, $1<s\le 8/7$, then every weak solution
 $(u,p)$ belongs to $W^{2,s}({\cal G})^3\times W^{1,s}({\cal G})$.
\end{itemize}

{\em The mixed problem with boundary conditions} (i)--(iii). Let
$(u,p) \in W^{1,2}({\cal G})^3\times L_2({\cal G})$ be a weak
solution of the problem (\ref{NS1}), (\ref{NS2}), where $g=0$,
$h_j=0$, $\phi_j=0$ for $j=1,\ldots,N$, and $d_k\le 2$ for all $k$
(i.e., the Neumann condition does not appear in the boundary
conditions). We assume that the Dirichlet condition is given on at
least one of the adjoining faces of every edge. Then, by
\cite[Th.6.1.5]{kmr2}, the strip $-1\le \mbox{Re}\, \lambda\le 0$
is free of eigenvalues of the pencils ${\mathfrak A}_j(\lambda)$.
Furthermore, we have $\mu_k> 1/2$ if the Dirichlet condition is
given on both adjoining faces of the edge $M_k$. For the other
indices $k$, we have $\mu_k > 1/4$ and $\mu_k >1/3$ if $\theta_k <
\frac 32 \pi$.
\begin{itemize}
\item If $f\in (W^{1,s'}({\cal G})^*)^3$, $2< s \le 8/3$,
  then $(u,p)\in W^{1,s}({\cal G})^3\times L_s({\cal G})$. Suppose that $\theta_k<\frac 32 \pi$ if
  the boundary condition (ii) or (iii) is given on one of the adjoining faces of the edge $M_k$.
  Then this result is even true for $2<s \le 3$.
\item If $f\in (W^{1,2}({\cal G})^*)^3\cap L_s({\cal G})^3$, $1<s\le 8/7$, then $(u,p) \in W^{2,s}({\cal G})^3
  \times W^{1,s}({\cal G})$. Suppose that $\theta_k  < 3\, \mbox{arccos}\frac 14$ if the Dirichlet condition
  is given on both adjoining faces of $M_k$, $\theta_k < \frac 32 \, \mbox{arccos}\frac 14$
  if the boundary condition (ii) is given on one of the
  adjoining faces of $M_k$, and $\theta_k < \frac 34 \pi$ if the boundary condition (iii) is given
  on one of the adjoining faces of $M_k$. Then the last result is true for $1< s\le 3/2$.
\end{itemize}
Note that in the last case, we have $\mu_k >2/3$ for
$k=1,\ldots,m$.

Finally, we assume that the homogeneous Dirichlet condition $u=0$
is given on the faces $\Gamma_1,\ldots,\Gamma_{N-1}$, while the
homogeneous boundary condition (iii) is given on $\Gamma_N$. Let
$I$ be the set of all $k$ such that $M_k \subset \bar{\Gamma}_N$
and $I'=\{1,\ldots,m\}\backslash I.$ We suppose that the
polyhedron ${\cal G}$ is convex and $\theta_k < \pi/2$ for $k\in
I$. Then $\mu_k>1$ for all $k$, and the strip $-1/2\le \mbox{Re}\,
\lambda \le 1$ contains only the simple eigenvalue $\lambda=1$ of
the pencils ${\mathfrak A}_j(\lambda)$ (see
\cite[Th.6.2.7]{kmr2}). If $\theta_k< \frac 38 \pi$ for $k\in I$
and $\theta_k< \frac 34 \pi$ for $k\in I'$, then even $\mu_k>4/3$.
This implies the following result.
\begin{itemize}
\item Let $f\in (W^{1,2}({\cal G})^*)^3\cap L_s({\cal G})^3$, $1<s\le 2$.
  Then any weak solution belongs to $W^{2,s}({\cal G})^3\times W^{1,s}({\cal G})$. If
  $\theta_k< \frac 38 \pi$ for $k\in I$ and $\theta_k< \frac 34 \pi$ for $k\in I'$, then the result
  holds even for $1<s<3$.
\end{itemize}
Furthermore analogously to the Dirichlet problem, the following
assertion holds.
\begin{itemize}
\item Let $f\in (W^{1,2}({\cal G})^*)^3\cap C^{-1,\sigma}({\cal G})$. Then for sufficiently small $\sigma$,
  we have $(u,p) \in C^{1,\sigma}({\cal G}) \times C^{0,\sigma}({\cal G})$.
\end{itemize}

\setcounter{equation}{0}
\section{Existence of weak solutions in $W^{1,s}({\cal G})\times L_s({\cal G})$, $s<2$}

In Section 1 we proved the existence of weak solutions of the
boundary value problem in $W^{1,2}({\cal G})\times L_2({\cal G})$.
Using the regularity result of Theorem \ref{dt1}, we obtain also
the existence of weak solutions in $W^{1,s}({\cal G})\times
L_s({\cal G})$ for sufficiently small $s>2$. In this section, we
will prove that weak solutions exist also in the space
$W^{1,s}({\cal G})\times L_s({\cal G})$ with $s<2$ provided the
norms of the right-hand sides of (\ref{a1}), (\ref{a2}) in the
corresponding Sobolev spaces are sufficiently small. Throughout
this section, we suppose that the Dirichlet condition is given on
at least one of the adjoining faces of every edge $M_k$.

\subsection{Solvability of the linearized problem in a cone}

Let ${\cal K}$ be the same polyhedral cone as in Section 2.1. We
consider weak solutions $(u,p) \in V_{\beta,\delta}^{1,s}({\cal
G})^3 \times V_{\beta,\delta}^{0,s}({\cal G})$ of the linear
Stokes system in ${\cal K}$ with boundary conditions (i)--(iv) on
the faces $\Gamma_j$. This means that $(u,p)$ satisfies the
equations
\begin{eqnarray} \label{StokesK1}
&& b_{\cal K}(u,v) - \int_{\cal K} p\, \nabla\cdot v\, dx =
F(v)\quad \mbox{for all }
    v\in V_{-\beta,-\delta}^{1,s'}({\cal K})^3,\  S_j v|_{\Gamma_j}=0, \\ \label{StokesK2}
&& - \nabla \cdot u = g\ \ \mbox{in }{\cal K}, \qquad S_j u = h_j
\ \mbox{ on }\Gamma_j,\ j=1,\ldots,N.
\end{eqnarray}
Here $b_{\cal K}$ denotes the bilinear form (\ref{a3}), where
${\cal G}$ has to be replaced by ${\cal K}$. We define the space
$V_{\beta,\delta}^{-1,s}({\cal K};S)$ as the set of all linear and
continuous functionals on the space $\{  v\in
V_{-\beta,-\delta}^{1,s'}({\cal K})^3:\  S_j v|_{\Gamma_j}=0\}$,
where $s'=s/(s-1)$. Furthermore, the pencils $A_k(\lambda)$ for
the edges $M_k$ and ${\mathfrak A}(\lambda)$ for the vertex of the
cone ${\cal K}$ are defined as in Section 3.1. If the Dirichlet
condition (i) is given on at least one of the adjoining faces of
the edge $M_k$, then $\lambda=0$ is not an eigenvalue of the
pencil $A_k(\lambda)$.

The following lemma is proved in \cite{mr-04b} under the condition
$\max(1-\mbox{Re}\,\lambda_1^{(k)},0)< \delta_k+2/s<1$. Using the
sharper estimates of Green's matrix given in \cite{mr-04a} for the
case when $\lambda=0$ is not an eigenvalue of the pencils
$A_k(\lambda)$, this theorem can be proved in the same way if
\begin{equation} \label{z1}
1-\mbox{Re}\,\lambda_1^{(k)}<\delta_k+2/s<1+\mbox{Re}\,\lambda_1^{(k)}
\end{equation}
for $k=1,\ldots,N.$

\begin{Le} \label{zl1}
Suppose that $F \in V_{\beta,\delta}^{-1,s}({\cal K};S)$, $g\in
V_{\beta,\delta}^{0,s}({\cal K})$, $h_j \in
V_{\beta,\delta}^{1-1/s,s}(\Gamma_j)$, there are no eigenvalues of
the pencil ${\mathfrak A}(\lambda)$ on the line $\mbox{\em
Re}\lambda = 1-\beta_j-3/s$, and the components of $\delta$
satisfy the inequalities {\em (\ref{z1})}. Then there exists a
unique solution $(u,p)\in V_{\beta,\delta}^{1,s}({\cal K})^3\times
V_{\beta,\delta}^{0,s}({\cal K})$ of problem {\em
(\ref{StokesK1}), (\ref{StokesK2})}.
\end{Le}

Moreover, the following regularity results hold analogously to
\cite[Le.4.4 and Th.4.4]{mr-04b}.

\begin{Le} \label{zl2}
{\em 1)} Suppose that in addition to the assumptions of Lemma {\em
\ref{zl1}}, we have $F \in V_{\beta',\delta'}^{-1,t}({\cal K};S)$,
$g\in V_{\beta',\delta'}^{0,t}({\cal K})$, $h_j \in
V_{\beta',\delta'}^{1-1/t,t}(\Gamma_j)$, where $1-\mbox{\em
Re}\,\lambda_1^{(k)}<\delta'_k+2/t<1+\mbox{\em
Re}\,\lambda_1^{(k)}$ for $k=1,\ldots,N$ and $\beta'$ is such that
there are no eigenvalues of the pencil ${\mathfrak A}(\lambda)$ in
the closed strip between the lines $\mbox{\em Re}\lambda =
1-\beta-3/s$ and $\mbox{\em Re}\lambda = 1-\beta'-3/t$. Then the
solution $(u,p)\in V_{\beta,\delta}^{1,s}({\cal K})^3\times
V_{\beta,\delta}^{0,s}({\cal K})$ belongs to
$V_{\beta',\delta'}^{1,t}({\cal K})^3\times
V_{\beta',\delta'}^{0,t}({\cal K})$.

{\em 2)} Suppose that in addition to the assumptions of Lemma {\em
\ref{zl1}}, $g\in V_{\beta',\delta'}^{1,t}({\cal K})$, $h_j \in
V_{\beta',\delta'}^{2-1/t,t}(\Gamma_j)$, and the functional $F$
has the form
\[
F(v) = \int_{\cal K} f\cdot v\, dx + \sum_{j=1}^N \Phi_j\cdot v\,
dx
\]
with vector function $f\in V_{\beta',\delta'}^{0,t}({\cal K})$,
$\Phi_j \in V_{\beta',\delta'}^{1-1/t,t}(\Gamma_j)$, where
$\beta'$ is such that there are no eigenvalues of the pencil
${\mathfrak A}(\lambda)$ in the closed strip between the lines
$\mbox{\em Re}\lambda = 1-\beta-3/s$ and $\mbox{\em Re}\lambda =
2-\beta'-3/t$, and the components of $\delta'$ satisfy the
inequalities $2-\mbox{\em Re}\,\lambda_1^{(k)}<\delta'_k+2/t <
2+\mbox{\em Re}\,\lambda_1^{(k)}$. Then the solution $(u,p)\in
V_{\beta,\delta}^{1,s}({\cal K})^3 \times
V_{\beta,\delta}^{0,s}({\cal K})$ belongs to
$V_{\beta',\delta'}^{2,t}({\cal K})^3\times
V_{\beta',\delta'}^{1,t}({\cal K})$.
\end{Le}

\subsection{Solvability of the linearized problem in ${\cal G}$}

We consider the operator
\begin{equation} \label{z2}
V_{\beta,\delta}^{1,s}({\cal G})^3\times
V_{\beta,\delta}^{0,s}({\cal G}) \ni (u,p) \to
  (F,g,h) \in V_{\beta,\delta}^{-1,s}({\cal G};S)
  \times V_{\beta,\delta}^{0,s}({\cal G})\times \prod_j V_{\beta,\delta}^{1-1/s,s}(\Gamma_j)
\end{equation}
of problem (\ref{Stokes1}), (\ref{Stokes2}) and denote this
operator by ${\cal A}_{s,\beta,\delta}$. Here again
$V_{\beta,\delta}^{-1,s}({\cal G};S)$ is defined as the dual space
of $\{  v\in V_{-\beta,-\delta}^{1,s'}({\cal G})^3:\  S_j
v|_{\Gamma_j}=0\}$. We show that this operator is Fredholm if
there are no eigenvalues of the pencil ${\mathfrak A}_j(\lambda)$
on the line $\mbox{Re}\lambda = 1-\beta_j-3/s$, $j=1,\ldots,d$,
and the components of $\delta$ satisfy the inequalities
\begin{equation} \label{z3}
1- \inf_{\xi\in M_k}
\mbox{Re}\,\lambda_1(\xi)<\delta_k+2/s<1+\inf_{\xi\in
M_k}\mbox{Re}\,\lambda_1(\xi)
\end{equation}
for $k=1,\ldots,m$. For this end, we construct a left and right
regularizer for the operator ${\cal A}_{s,\beta,\delta}$.

\begin{Le} \label{zl3}
Let ${\cal U}$ be a sufficiently small open subset of ${\cal G}$
and let $\varphi$ be a smooth function with support in $\bar{\cal
U}$. Suppose that there are no eigenvalues of the pencil
${\mathfrak A}_j(\lambda)$ on the line $\mbox{\em Re}\lambda =
1-\beta_j-3/s$, $j=1,\ldots,d$, and the components of $\delta$
satisfy {\em (\ref{z3})}. Then there exists an operator ${\cal R}$
continuously mapping the space of all $(F,g,h) \in
V_{\beta,\delta}^{-1,s}({\cal G};S) \times
V_{\beta,\delta}^{0,s}({\cal K})\times \prod_j
V_{\beta,\delta}^{1-1/s,s}(\Gamma_j)$ with support in $\bar{\cal
U}$ onto $V_{\beta,\delta}^{1,s}({\cal G})^3\times
V_{\beta,\delta}^{0,s}({\cal G})$ such that $\varphi {\cal
A}_{s,\beta,\delta}{\cal R}(F,g,h)= \varphi(F,g,h)$ for all
$(F,g,h)$ with support in $\bar{\cal U}$ and ${\cal R}{\cal
A}_{s,\beta,\delta}(u,p)=(u,p)$ for all $(u,p)$ with support in
$\bar{\cal U}$.
\end{Le}

P r o o f. Suppose first that $\bar{\cal U}$ contains the vertex
$x^{(1)}$ of ${\cal G}$. Then there exists a diffeomorphism
$\kappa$ mapping ${\cal U}$ onto a subset ${\cal V}$ of a
polyhedral cone ${\cal K}$ with vertex at the origin such that
$\kappa(x^{(1)})=0$ and the Jacobian matrix $\kappa'$ coincides
with the identity matrix $I$ at $x^{(1)}$. We assume that the
supports of $u$ and $p$ are contained in $\bar{\cal U}$. Then the
coordinate change $y=\kappa(x)$ transforms (\ref{Stokes1}),
(\ref{Stokes2}) into
\begin{eqnarray} \label{St1}
&& \tilde{b}(\tilde{u},\tilde{v})- \int_{\cal K} \tilde{p}\,
\tilde{\cal D}\tilde{v}\,
  |\mbox{det}\, \kappa'|^{-1}\, dy =\tilde{F}(\tilde{v}) \ \mbox{ for all } \tilde{v}\in
  V_{\beta,\delta}^{1,s'}({\cal K})^3, \ S_j\tilde{v}=0 \mbox{ on }\Gamma_j^\circ, \\ \label{St2}
&& -\tilde{\cal D}\tilde{u}=\tilde{g}\ \mbox{ in }{\cal K}, \quad
S_j\tilde{u}=\tilde{h}_j \ \mbox{ on }\Gamma_j^\circ,
\end{eqnarray}
where $\tilde{u} = u \circ \kappa^{-1}$,
$\tilde{F}(\tilde{v})=F(\tilde{v}\circ\kappa)$, $\Gamma_j^\circ$
are the faces of ${\cal K}$, $\tilde{\cal D}$ is a first order
differential operator of the form
\[
\tilde{\cal D} \tilde{u} = \big( D(y)\nabla_y\big)\cdot \tilde{u},
\]
and $\tilde{b}$ is a bilinear form having the representation
\[
\tilde{b}(\tilde{u},\tilde{v}) = 2\nu\, \int_{\cal K}
\sum_{i,j=1}^3 B_{i,j}(y)\, \partial_{y_i}\tilde u \cdot
  \partial_{y_j}\tilde{v}\, dy.
\]
Here $D(y)$ and $B_{i,j}(y)$ are quadratic matrices such that
$D(0)=I$ and
\[
\sum_{i,j=1}^3 B_{i,j}(0)\, \partial_{y_i}\tilde u \cdot
\partial_{y_j}\tilde{v} =
  \sum_{i,j=1}^3 \varepsilon_{i,j}(\tilde{u})\, \varepsilon_{i,j}(\tilde{v}).
\]
Let $\zeta$ be an infinitely differentiable cut-off function on
$[0,\infty)$ equal to 1 in $[0,1)$ and to zero in $(2,\infty)$.
For arbitrary positive $\epsilon$, we define
$\zeta_\epsilon(y)=\zeta(|y|/\epsilon)$, Moreover, we put
$\zeta_0=0$ and $\eta_\epsilon=1-\zeta_\epsilon$ for $\epsilon\ge
0$. We consider the operator
\begin{equation} \label{z4}
V_{\beta,\delta}^{1,s}({\cal K})^3\times
V_{\beta,\delta}^{0,s}({\cal K}) \ni (\tilde{u},\tilde{p}) \to
  (\tilde{F},\tilde{g},\tilde{h}) \in V_{\beta,\delta}^{-1,s}({\cal K};S)
  \times V_{\beta,\delta}^{0,s}({\cal K})\times \prod_j V_{\beta,\delta}^{1-1/s,s}(\Gamma_j^\circ)
\end{equation}
defined by
\begin{eqnarray*}
&& b_\epsilon(\tilde{u},\tilde{v}) - \int_{\cal K} \tilde{p} \,
\big( \zeta_\epsilon\, \tilde{\cal D}
  \tilde{v}\, |\mbox{det}\, \kappa'|^{-1} + \eta_\epsilon\, \nabla_y\cdot\tilde{v}\big)\, dy
  = \tilde{F}(\tilde{v}) \ \mbox{ for all } \tilde{v}\in V_{\beta,\delta}^{1,s'}({\cal K})^3 , \
  S_j\tilde{v}=0 \mbox{ on }\Gamma_j^\circ,\\
&& -\big( \zeta_\epsilon
D(y)\nabla_y+\eta_\epsilon\nabla_y\big)\cdot\tilde{u}=\tilde{g} \
  \mbox{ in }{\cal K}, \quad S_j\tilde{u}=\tilde{h}_j \ \mbox{ on }\Gamma_j^\circ,
\end{eqnarray*}
where
\[
b_\epsilon(\tilde{u},\tilde{v}) = 2\nu \int_{\cal K}
\sum_{i,j=1}^3 \big( \zeta_\epsilon B_{i,j}(y)
  + \eta_\epsilon B_{i,j}(0)\big)\, \partial_{y_i}\tilde u \cdot  \partial_{y_j}\tilde{v}\, dy.
\]
We denote the operator (\ref{z4}) by $\tilde{\cal A}_\epsilon$.
According to Lemma \ref{zl1}, the operator $\tilde{\cal A}_0$ is
an isomorhism. Since the norm of $\tilde{\cal A}_0-\tilde{\cal
A}_\epsilon$ is small for small $\epsilon$, the operator
$\tilde{\cal A}_\epsilon$ is an isomorphism if
$\epsilon\le\epsilon_0$ and $\epsilon_0$ is sufficiently small. We
may assume that $\zeta_\epsilon=1$ on ${\cal V}$ for
$\epsilon=\epsilon_0$. Then problem (\ref{St1}), (\ref{St2}) can
be written as $\tilde{\cal A}_{\epsilon_0}(\tilde{u},\tilde{p}) =
(\tilde{F},\tilde{g},\tilde{h})$ if the supports of $\tilde{u}$
and $\tilde{p}$ are contained in $\bar{\cal V}$. Let
\begin{equation} \label{z5}
u(x) = \tilde{u}(\kappa(x)),\ p(x) = \tilde{p}(\kappa(x))\ \mbox{
for }x\in {\cal U},\quad\mbox{where }
  (\tilde{u},\tilde{p}) = \tilde{\cal A}_{\epsilon_0}^{-1}\, (\tilde{F},\tilde{g},\tilde{h}).
\end{equation}
Outside ${\cal U}$, let $(u,p)$ be continuously extended to a
vector function from $V_{\beta,\delta}^{1,s}({\cal G})^3\times
V_{\beta,\delta}^{0,s}({\cal G})$. The so defined mapping
$(F,g,h)\to (u,p)$ is denoted by ${\cal R}$ and has the desired
properties.

Suppose now that $\bar{\cal U}$ contains an edge point $\xi\in
M_1$ but no points of other edges and no vertices of ${\cal G}$.
Then again there exists a diffeomorphism mapping ${\cal U}$ onto a
subset of a cone ${\cal K}$. We assume that the point
$\kappa(\xi)$ lies on the edge $M_1^\circ$ of ${\cal K}$ and
coincides with the origin (in contrast to the first part of the
proof, the vertex of the cone is not the origin). Let $\tilde{\cal
A}_\epsilon$ be the same operator as above. Then there exist a
number $\beta_0$ and a tuple $\delta'$, $\delta'_1=\delta_1$, such
that $\tilde{\cal A}_0$ and for sufficiently small $\epsilon$ also
$\tilde{\cal A}_\epsilon$ are isomorphisms
\[
V_{\beta_0,\delta'}^{1,s}({\cal K})^3\times
V_{\beta_0,\delta'}^{0,s}({\cal K}) \to
  V_{\beta_0,\delta'}^{-1,s}({\cal K};S) \times V_{\beta_0,\delta'}^{0,s}({\cal K})
  \times \prod_j V_{\beta_0,\delta'}^{1-1/s,s}(\Gamma_j^\circ).
\]
Since $\bar{\cal U}$ does not contain points of the edges $M_k$,
$k\not=1$, the vector function (\ref{z5}) can be continuously
extended to a vector function from $V_{\beta,\delta}^{1,s}({\cal
G})^3\times V_{\beta,\delta}^{0,s}({\cal G})$. The so defined
mapping $(F,g,h) \to (u,p)$ defines the desired operator ${\cal
R}$. Analogously, the lemma can be proved for the case when
$\bar{\cal U}$ contains no edge points of ${\cal G}$. \hfill
$\Box$

\begin{Rem} \label{zr1}
{\em Suppose that
\[
2- \inf_{\xi\in M_k}
\mbox{Re}\,\lambda_1(\xi)<\delta'_k+2/t<2+\inf_{\xi\in
M_k}\mbox{Re}\,\lambda_1(\xi)
  \quad\mbox{for }k=1,\ldots,m
\]
and there are no eigenvalues of the pencil ${\mathfrak
A}_j(\lambda)$ in the closed strip between the lines
$\mbox{Re}\lambda = 1-\beta_j-3/s$ and $\mbox{Re}\lambda =
2-\beta'_j-3/t$. Then it follows from Lemma \ref{zl2} that the
operator ${\cal R}$ constructed in the proof of Lemma \ref{zl3}
continuously maps the subspace of all $(F,g,h)$, where
\[
g\in V_{\beta,\delta}^{0,s}({\cal G}) \cap
V_{\beta',\delta'}^{1,t}({\cal G}), \quad
  h_j \in V_{\beta,\delta}^{1-1/s,s}(\Gamma_j)\cap V_{\beta',\delta'}^{2-1/t,t}(\Gamma_j)
\]
and the functional $F\in V_{\beta,\delta}^{-1,s}({\cal G};S)$ has
the form
\[
F(v)= \int_{\cal G} f\cdot v\, dx + \sum_j \int_{\Gamma_j}
\Phi_j\cdot v\, dx\quad\mbox{with vector functions }
  f\in V_{\beta',\delta'}^{0,t}({\cal G})^3,\ \Phi_j \in V_{\beta',\delta'}^{1-1/t,t}(\Gamma_j)^3,
\]
into $V_{\beta',\delta'}^{2,t}({\cal G})^3 \times
V_{\beta',\delta'}^{1,t}({\cal G})$.}
\end{Rem}

\begin{Le} \label{zl4}
Suppose that there are no eigenvalues of the pencil ${\mathfrak
A}_j(\lambda)$ on the line $\mbox{\em Re}\lambda = 1-\beta_j-3/s$,
$j=1,\ldots,d$, and the components of $\delta$ satisfy {\em
(\ref{z3})}. Then there exists a continuous operator
\[
{\cal R}: \, V_{\beta,\delta}^{-1,s}({\cal G};S)\times
V_{\beta,\delta}^{0,s}({\cal K})\times
  \prod_j V_{\beta,\delta}^{1-1/s,s}(\Gamma_j) \to V_{\beta,\delta}^{1,s}({\cal G})^3
  \times V_{\beta,\delta}^{0,s}({\cal G})
\]
such that ${\cal R}{\cal A}_{s,\beta,\delta}-I$ and ${\cal
A}_{s,\beta,\delta}{\cal R}-I$ are compact operators in
$V_{\beta,\delta}^{1,s}({\cal G})^3 \times
V_{\beta,\delta}^{0,s}({\cal G})$ and
$V_{\beta,\delta}^{-1,s}({\cal G};S)\times
V_{\beta,\delta}^{0,s}({\cal G})\times
  \prod_j V_{\beta,\delta}^{1-1/s,s}(\Gamma_j)$, respectively.
\end{Le}

P r o o f. For the sake of brevity, we write ${\cal A}$ instead of
${\cal A}_{s,\beta,\delta}$. Let $\{ {\cal U}_j\}$ be a
sufficiently fine open covering of ${\cal G}$, and let
$\varphi_j$, $\psi_j$ be infinitely differentiable functions such
that
\[
\mbox{supp}\, \varphi_j \subset \mbox{supp}\, \psi_j\subset {\cal
U}_j, \quad \varphi_j\, \psi_j = \varphi_j,
  \quad\mbox{and}\quad \sum_j \varphi_j=1.
\]
For every $j$, there exists an operator ${\cal R}_j$ having the
properties of Lemma \ref{zl2} for ${\cal U}={\cal U}_j \cap {\cal
G}$. We consider the operator ${\cal R}$ defined by
\[
{\cal R}\, (F,g,h) = \sum_j \varphi_j \, {\cal R}_j\, \psi_j\,
(F,g,h).
\]
Obviously,
\[
{\cal R}\, {\cal A} (u,p) = \sum_j \varphi_j {\cal R}_j\, \big(
{\cal A}\psi_j(u,p)-[{\cal A},\psi_j]\, (u,p)\big)
  = (u,p) - \sum_j \varphi_j\, {\cal R}_j\, [{\cal A},\psi_j]\, (u,p),
\]
where $[{\cal A},\psi_j]$ is the commutator of ${\cal A}$ and
$\psi_j$. Here, the mapping $(u,p) \to [{\cal A},\psi_j]\, (u,p)$
is continuous from $V_{\beta,\delta}^{1,s}({\cal G})^3 \times
V_{\beta,\delta}^{0,s}({\cal G})$ into the set of all $(F,g,h)$,
where $g\in V_{\beta',\delta'}^{1,s}({\cal G})$, $h=0$, and $F$ is
a functional of the form
\[
F(v)= \int_{\cal G} f\cdot v\, dx + \sum_j \int_{\Gamma_j}
\Phi_j\cdot v\, dx,\quad\mbox{where }
  f\in V_{\beta',\delta'}^{0,s}({\cal G})^3,\ \Phi_j \in V_{\beta',\delta'}^{1-1/s,s}(\Gamma_j)^3,
\]
with arbitrary $\beta'\ge \beta$, $\delta'\ge \delta$. We can
choose $\beta'$ and $\delta'$ such that $\beta_j\le \beta'_j<
\beta_j+1$ for $j=1,\ldots,d$, $\delta_k\le \delta'_k<\delta_k+1$
for $k=1,\ldots,m$, and $\beta',\delta'$ satisfy the conditions of
Remark \ref{zr1} with $t=s$. Then the mapping $(u,p) \to {\cal
R}_j\, [{\cal A},\psi_j]\, (u,p)$ is continuous from
$V_{\beta,\delta}^{1,s}({\cal G})^3 \times
V_{\beta,\delta}^{0,s}({\cal G})$ into
$V_{\beta',\delta'}^{2,s}({\cal G})^3 \times
V_{\beta',\delta'}^{1,s}({\cal G})$. Since the last space is
compactly imbedded into $V_{\beta,\delta}^{1,s}({\cal G})^3 \times
V_{\beta,\delta}^{0,s}({\cal G})$ for $\beta'_j<\beta_j+1$,
$\delta'_k<\delta_k+1$ (cf. \cite[Le.6.2.1]{kmr1}), it follows
that ${\cal R} {\cal A} -I$ is compact.
Analogously, the compactness of ${\cal A} {\cal R} -I$ holds. \hfill $\Box$ \\

From Lemma \ref{zl4} it follows that the operator ${\cal
A}_{s,\beta,\delta}$ is Fredholm under the conditions of this
lemma. Using Theorem \ref{Stokest1} and Lemma \ref{zl2}, we can
prove the following theorem on the existence of unique solutions.

\begin{Th} \label{zt1}
Let $F\in V_{\beta,\delta}^{-1,s}({\cal G};S)$, $g\in
V_{\beta,\delta}^{0,s}({\cal G})$, and $h_j \in
V_{\beta,\delta}^{1-1/s,s}(\Gamma_j)^{3-d_j}$. We suppose that the
components of $\delta$ satisfy condition {\em (\ref{z1})} and that
there are no eigenvalues of the pencils ${\mathfrak A}_j
(\lambda)$ in the closed strip between the lines $\mbox{\em Re}\,
\lambda=-1/2$ and $\mbox{\em Re}\, \lambda=1-\beta-3/s$. In the
case when $d_j\in \{ 0,2\}$ for all $j$, we assume in addition
that $g$ and $h_j$ satisfy condition {\em (\ref{1Stokest1})}. Then
there exists a solution $(u,p)\in V_{\beta,\delta}^{1,s}({\cal
G})^3\times V_{\beta,\delta}^{0,s}({\cal G})$ of problem {\em
(\ref{Stokes1}), (\ref{Stokes2})}. (Here $V$ has to be replaced by
the set of all $v\in V_{-\beta,-\delta}^{1,s'}({\cal G})$
satisfying $S_j v=0$ on $\Gamma_j$.) The vector function $u$ is
unique, $p$ is unique if $d_j\not\in \{0,2\}$ for at least one
$j$, $p$ is unique up to a constant if $d_j \in \{0,2\}$ for all
$j$.
\end{Th}

P r o o f. First note that in the case when $d_j \in \{0,2\}$ for
all $j$, the spectra both of ${\mathfrak A}_j(\lambda)$ and
$A_k(\lambda)$ contain the eigenvalue $\lambda=1$. An eigenvector
corresponding to this eigenvalue is $(U,P)=(0,1)$. Furthermore,
the spectra of the pencils ${\mathfrak A}_j(\lambda)$ contain the
eigenvalue $\lambda=-2$. Therefore from the conditions of the
lemma on $\beta$ and $\delta$ it follows in particular that
$0<\beta_j+3/s<3$ and $0<\delta_k+2/s<2$. Then $L_\infty({\cal G})
\subset V_{\beta,\delta}^{0,s}({\cal G}) \subset L_1({\cal G})$.
In particular, any constant belongs to the space
$V_{\beta,\delta}^{0,s}({\cal G})$, and the integrals in
(\ref{1Stokest1}) exist for $g\in V_{\beta,\delta}^{0,s}({\cal
G})$, $h_j \in V_{\beta,\delta}^{1-1/s,s}(\Gamma_j)^{3-d_j}$.

We prove the uniqueness of $u$ and $p$. Let $(u,p) \in
V_{\beta,\delta}^{1,s}({\cal G})^3\times
V_{\beta,\delta}^{0,s}({\cal G})$ be a solution of problem
(\ref{Stokes1}), (\ref{Stokes2}) with $F=0$, $g=0$, $h_j=0$. Then
according to Lemma \ref{zl2}, we have $u\in V_{0,0}^{1,2}({\cal
G})^3 \subset W^{1,2}({\cal G})^3$ and $p\in L_2({\cal G})$. From
Theorem \ref{Stokest1} it follows that $u=0$ and $p$ is constant,
$p=0$ if $d_j\in \{1,3\}$ for at least one $j$.

We prove the existence of solutions for the case when $d_j\in
\{0,2\}$ for all $j$. Due to Lemma \ref{zl4} and the uniqueness of
the solution, every solution $(u,p) \in
V_{\beta,\delta}^{1,s}({\cal G})^3\times
V_{\beta,\delta}^{0,s}({\cal G})$ of problem (\ref{Stokes1}),
(\ref{Stokes2}), $\int_{\cal G} p\, dx=1$, satisfies the
inequality
\begin{equation} \label{1zt1}
\| u\|_{V_{\beta,\delta}^{1,s}({\cal G})^3} + \|
p\|_{V_{\beta,\delta}^{0,s}({\cal G})}
  \le c\, \Big( \| F\|_{V_{\beta,\delta}^{-1,s}({\cal G};S))} + \| g\|_{V_{\beta,\delta}^{0,s}({\cal G})}
  + \sum_j \| h_j\|_{V_{\beta,\delta}^{1-1/s,s}(\Gamma_j)}\Big)
\end{equation}
with a constant $c$ independent of $u$ and $p$. Let $F \in
V_{\beta,\delta}^{-1,s}({\cal G};S)$, $g\in
V_{\beta,\delta}^{0,s}({\cal G})$ and $h_j \in
V_{\beta,\delta}^{1-1/s,s}(\Gamma_j)^{3-d_j}$ satisfying
(\ref{1Stokest1}) be given. Then there exist sequences
$\{F^{(n)}\} \subset V^* \cap V_{\beta,\delta}^{-1,s}({\cal
G};S)$, $\{ g^{(n)}\} \subset L_2({\cal G}) \cap
V_{\beta,\delta}^{0,s}({\cal G})$ and $\{ h_j^{(n)}\} \subset
W^{1/2,2}(\Gamma_j)^{3-d_j}\cap
V_{\beta,\delta}^{1-1/s,s}(\Gamma_j)^{3-d_j}$ converging to $F$,
$g$, and $h_j$, respectively. From (\ref{1Stokest1}) it follows
that the sequences of the numbers
\[
a_n = \int_{\cal G} g^{(n)}\, dx + \sum_{j:\, d_j=0}
\int_{\Gamma_j} h_j^{(n)}\cdot n\, dx +
  \sum_{j:\, d_j=2} \int_{\Gamma_j} h_j^{(n)}\, dx
\]
converges to zero. Therefore, the sequence of the functions
$\tilde{g}^{(n)} = g^{(n)}- \frac 1{|{\cal G}|}\, a_n$ converges
also to $g$. Moreover, $\tilde{g}^{(n)}$ and $h_j^{(n)}$ satisfy
condition (\ref{1Stokest1}). By Theorem \ref{Stokest1}, there
exist solutions $(u^{(n)},p^{(n)}) \in W^{1,2}({\cal G})^3\times
L_2({\cal G})$ of problem (\ref{Stokes1}), (\ref{Stokes2}) with
right-hand sides $F^{(n)}$, $\tilde{g}^{(n)}$ and $h_j^{(n)}$,
$\int_{\cal G} p^{(n)}\, dx =0$.  From Lemma \ref{zl2} and from
the imbedding $W^{1,2}({\cal G}) \subset
V_{0,\varepsilon}^{1,2}({\cal G})$ with arbitrary positive
$\varepsilon$ it follows that $(u^{(n)},p^{(n)}) \in
V_{\beta,\delta}^{1,s}({\cal G})^3 \times
V_{\beta,\delta}^{0,s}({\cal G})$. Due to (\ref{1zt1}), the vector
functions $(u^{(n)},p^{(n)})$ form a Cauchy sequence in
$V_{\beta,\delta}^{1,s}({\cal G})^3 \times
V_{\beta,\delta}^{0,s}({\cal G})$. The limit of this sequence
solves problem (\ref{Stokes1}), (\ref{Stokes2}). The proof of the
existence of solutions in the case when $d_j\in \{ 1,3\}$ for at
least one $j$ proceeds analogously. \hfill $\Box$

\subsection{Solvability of the nonlinear problem}

We consider the nonlinear problem (\ref{NS1}), (\ref{NS2}). For
the proof of the existence of solutions in
$V_{\beta,\delta}^{1,s}({\cal G})^3 \times
V_{\beta,\delta}^{0,s}({\cal G})$, we have to have to show
inequalities analogous to (\ref{Q1}), (\ref{Q2}) for the operator
\[
V_{\beta,\delta}^{1,s}({\cal G})^3 \ni u \to Qu =(u\cdot \nabla) u
\in V_{\beta,\delta}^{-1,s}({\cal G};S).
\]

\begin{Le} \label{zl5}
Suppose that $s>3/2$, $\beta_j+3/s\le 2$ for $j=1,\ldots,N$, and
$\delta_k+3/s\le 2$ for $k=1,\ldots,m$. Then
\[
\| u\, \partial_{x_j}v\|_{V_{\beta,\delta}^{-1,s}({\cal G})^3} \le
c\,
  \| u\|_{V_{\beta,\delta}^{1,s}({\cal G})} \ \| v\|_{V_{\beta,\delta}^{1,s}({\cal G})}
\]
for all $u,v \in V_{\beta,\delta}^{1,s}({\cal G})$, $j=1,2,3$.
\end{Le}

P r o o f. Let $t$ be a real number such that
\[
t\ge 3, \ t\ge s, \quad \mbox{and}\quad \frac 3s -1 \le \frac 3t <
3 - \frac 3s .
\]
For example, we can put $t=\max(s,3)$. Then according to Lemma
\ref{al8}, the space $V_{\beta,\delta}^{1,s}({\cal G})$ is
continuously imbedded into $V_{\beta',\delta'}^{0,t}({\cal G})$,
where $\beta'_j = \beta-1+3s^{-1} -  3t^{-1}$, $\delta'_k=
\delta_k-1 + 3s^{-1} - 3t^{-1}$. Let $q^{-1}= s^{-1}+t^{-1}$. From
the conditions on $t$ it follows that $q>1$. By H\"olders's
inequality, we have
\[
\| u\partial_{x_j}
v\|_{V_{\beta+\beta',\delta+\delta'}^{0,q}({\cal G})}
   \le c\, \| u\|_{V_{\beta',\delta'}^{0,t}({\cal G})} \ \| \partial_{x_j} v
   \|_{V_{\beta,\delta}^{0,s}({\cal G})}
\]
for arbitrary $u,v \in V_{\beta,\delta}^{1,s}({\cal G})$. Using
the continuity of the imbedding
$V_{\beta+\beta',\delta+\delta'}^{0,q}({\cal G})  \subset
V_{\beta,\delta}^{-1,s}({\cal G})$
which follows from Corollary \ref{ac1}, we obtain the desired inequality. \hfill $\Box$ \\

As a consequence of Lemma \ref{zl5}, the following inequalities
hold for arbitrary $u,v\in V_{\beta,\delta}^{1,s}({\cal G})^3$:
\begin{eqnarray} \label{Q3}
&& \| Qu\|_{V_{\beta,\delta}^{-1,s}({\cal G};S)} \le c\, \|
u\|^2_{V_{\beta,\delta}^{1,s}({\cal G})^3},\\ \label{Q4} && \|
Qu-Qv\|_{V_{\beta,\delta}^{-1,s}({\cal G};S)} \le c\, \big( \|
u\|_{V_{\beta,\delta}^{1,s}({\cal G})^3}
  + \| v\|_{V_{\beta,\delta}^{1,s}({\cal G})^3}\big) \,   \| u-v\|_{V_{\beta,\delta}^{1,s}({\cal G}))^3}.
\end{eqnarray}
Using these estimates, the following theorem can be proved
analogously to Theorem \ref{Stokest1a}.

\begin{Th} \label{zt2}
Let the conditions of Theorem {\em \ref{zt1}} on $F$, $g$ $h_j$,
$\beta$ and $\delta$ be satisfied. In addition, we assume that
$s>3/2$, $\beta_j+ 3/s\le 2$, $\delta_k+3/s\le 2$ and that
\[
\| F\|_{V_{\beta,\delta}^{-1,s}({\cal G};S)} + \|
g\|_{V_{\beta,\delta}^{0,s}({\cal G})}
 + \sum_j \| h_j \|_{V_{\beta,\delta}^{1-1/s,s}(\Gamma_j)^{3-d_j}}
\]
is sufficiently small. Then there exists a solution $(u,p) \in
V_{\beta,\delta}^{1,s}({\cal G})^3 \times
V_{\beta,\delta}^{0,s}({\cal G})$ of problem {\em (\ref{a1}),
(\ref{a2})}, where $V$ has to be replaced by the set of all $v\in
V_{-\beta,-\delta}^{1,s'}({\cal G})^3$ such that $S_jv=0$ on
$\Gamma_j$, $j=1,\ldots,N$. The function $u$ is unique on the set
of all functions with norm less than a certain positive
$\varepsilon$, $p$ is unique if $d_j \in \{ 1,3\}$ for at least
one $j$, otherwise $p$ is unique up to a constant.
\end{Th}

P r o o f. Let $(u^{(0)},p^{(0)})\in V_{\beta,\delta}^{1,s}({\cal
G})^3 \times V_{\beta,\delta}^{0,s}({\cal G})$. By Theorem
\ref{zt1}, the norm of $u^{(0)}$ can be assumed to be small.
Obviously, $(u,p)$ is a solution of problem (\ref{a1}), (\ref{a2})
if and only if $(w,q)=(u-u^{(0)},p-p^{(0)})$ is a solution of the
problem
\begin{eqnarray*}
&& b(w,v) - \int_{\cal G} q\, \nabla\cdot v\, dx = -\int_{\cal G}
Q(w+u^{(0)})\cdot v\, dx
  \ \mbox{ for all }v\in v\in V_{-\beta,-\delta}^{1,s'}({\cal G})^3,\ S_jv|_{\Gamma_j}=0, \\
&& \nabla\cdot w = 0 \ \mbox{ in }{\cal G}, \quad S_j u=0\ \mbox{
on }\Gamma_j,\ j=1,\ldots,N.
\end{eqnarray*}
This problem can be written as
\[
(w,q) = - AQ(w+u^{(0)})
\]
where $A$ is the inverse to the operator $F \to {\cal
A}_{s,\beta,\delta}(F,0,0)$ considered in the preceding
subsection. Due to (\ref{Q4}), the operator $(w,q) \to  -
AQ(w+u^{(0)})$ is contractive on the set of all $(w,q)\in
V_{\beta,\delta}^{1,s}({\cal G})^3 \times
V_{\beta,\delta}^{0,s}({\cal G})$ with norm $\le \varepsilon$ if
$\varepsilon$ and the norm of $u^{(0)}$ are sufficiently
small. Hence this operator has a fixed point. This proves the theorem. \hfill $\Box$\\

Finally, we give a result in nonweighted Sobolev spaces which
follows immediately from the last theorem.
\begin{itemize}
\item[] {\em  Let ${\cal G}$ be a polyhedron. We assume that on every face $\Gamma_j$, one of
  the boundary conditions} (i)--(iii) {\em is given, that the Dirichlet condition is given on at least
  one of the adjoining faces of every edge $M_k$, and that $\theta_k \le 3\pi/2$ if the boundary conditions}
  (ii) {\em or} (iii) {\em are given on one of the adjoining faces of $M_k$. Here, $\theta_k$ denotes again
  the angle   at the edge $M_k$. Then the problem
\begin{eqnarray*}
&& b(u,v) + \int_{\cal G} \sum_{j=1}^3 u_j\, \frac{\partial
u}{\partial x_j}\cdot v\, dx
  - \int_{\cal G} p\, \nabla\cdot v\, dx = F(v)\ \mbox{ for all }v\in W^{1,s'}({\cal G})^3,\
  S_jv=0\mbox{ on }\Gamma_j,\\
&& -\nabla \cdot u =0 \  \mbox{ in } {\cal G},\quad S_j u = 0\
\mbox{ on }\Gamma_j,\ j=1,\ldots,N,
\end{eqnarray*}
with $F\in W^{-1,s}({\cal G};S)$, $3/2 < s < 3$, has a solution
$(u,p)\in W^{1,s}({\cal G})^3 \times L_s({\cal G})$ if the norm of
$F$ is sufficiently small.}
\end{itemize}
For the proof it suffices to note that the strip $-1\le
\mbox{Re}\, \lambda\le 0$ does not contain eigenvalues of the
pencils ${\mathfrak A}_j(\lambda)$ and that $\mbox{Re}\,
\lambda_1^{(k)}\ge 1/3$ for $k=1,\ldots,m$. Therefore, the
conditions of Theorem \ref{zt2} are satisfied for $\beta=0$,
$\delta=0$, $3/2<s<3$.

\end{document}